# ThermoSim: Deep Learning based Framework for Modeling and Simulation of Thermal-aware Resource Management for Cloud Computing Environments

Sukhpal Singh Gill[1,11], Shreshth Tuli[2], Adel Nadjaran Toosi[3], Felix Cuadrado[1,4], Peter Garraghan[5], Rami Bahsoon[6], Hanan Lutfiyya[7], Rizos Sakellariou[8], Omer Rana[9], Schahram Dustdar[10] and Rajkumar Buyya[11]

[1]School of Electronic Engineering and Computer Science, Queen Mary University of London, UK
[2]Department of Computer Science and Engineering, Indian Institute of Technology (IIT), Delhi, India
[3]Faculty of Information Technology, Monash University, Clayton, Australia
[4]Technical University of Madrid (UPM), Spain
[5]School of Computing and Communications, Lancaster University, UK
[6]School of Computer Science, University of Birmingham, Birmingham, UK
[7]Department of Computer Science, University of Western Ontario, London, Canada
[8]School of Computer Science, University of Manchester, Oxford Road, Manchester, UK
[9]School of Computer Science and Informatics, Cardiff University, Cardiff, UK
[10]Distributed Systems Group, Vienna University of Technology, Vienna, Austria
[11]Cloud Computing and Distributed Systems (CLOUDS) Laboratory, School of Computing and Information Systems, The University of Melbourne, Australia

s.s.gill@qmul.ac.uk, shreshth.cs116@cse.iitd.ac.in, adel.n.toosi@monash.edu, felix.cuadrado@upm.es, p.garraghan@lancaster.ac.uk, r.bahsoon@cs.bham.ac.uk, hanan@csd.uwo.ca, rizos@manchester.ac.uk, ranaof@cardiff.ac.uk, dustdar@dsg.tuwien.ac.at, rbuyya@unimelb.edu.au

**Abstract.** Current cloud computing frameworks host millions of physical servers that utilize cloud computing resources in the form of different virtual machines. Cloud Data Center (CDC) infrastructures require significant amounts of energy to deliver large scale computational services. Moreover, computing nodes generate large volumes of heat, requiring cooling units in turn to eliminate the effect of this heat. Thus, overall energy consumption of the CDC increases tremendously for servers as well as for cooling units. However, current workload allocation policies do not take into account effect on temperature and it is challenging to simulate the thermal behaviour of CDCs. There is a need for a thermal-aware framework to simulate and model the behaviour of nodes and measure the important performance parameters which can be affected by its temperature. In this paper, we propose a lightweight framework, ThermoSim, for modelling and simulation of thermal-aware resource management for cloud computing environments. This work presents a Recurrent Neural Network based deep learning temperature predictor for CDCs which is utilized by ThermoSim for lightweight resource management in constrained cloud environments. ThermoSim extends the CloudSim toolkit helping to analyse the performance of various key parameters such as energy consumption, service level agreement violation rate, number of virtual machine migrations and temperature during the management of cloud resources for execution of workloads. Further, different energy-aware and thermal-aware resource management techniques are tested using the proposed ThermoSim framework in order to validate it against the existing framework (Thas). The experimental results demonstrate the proposed framework is capable of modelling and simulating the thermal behaviour of a CDC and ThermoSim framework is better than Thas in terms of energy consumption, cost, time, memory usage and prediction accuracy.

## 1 Introduction

Resource management is critical in cloud environment in which resource utilization, power consumption of servers, and storage play important roles. Provisioning and scheduling cloud resources is often based on availability, without considering other crucial parameters such as resource utilization or the server's thermal characteristics [1]. To realize this, a thermal-aware simulator for resource allocation mechanism is required [2]. The problem of allocating user workloads to a set of Physical Machines (PMs) or Virtual Machines (VMs) and allocating VMs on different server farms adhering to the terms of service as cited in Service Level Agreements (SLAs) and sustaining the Quality of Service (QoS) is stated as the service provisioning issue. Thus, cloud providers focus on developing energy-efficient approaches and policies [4].

Thermo-awareness in cloud refers to the consideration of thermal properties, such as thermal temperature of the host, CPU temperature, heat tolerance and thresholds, energy source (i.e. non-renewable vs. renewable), cooling considerations and mechanisms, cost etc. when dynamically managing cloud resources, scheduling and allocating workloads [20]. The explicit consideration of these properties can transform the way the cloud is managed and resources/PMs/VMs are dynamically allocated, leading to more energy-efficient computing and reduced carbon footprint [24] [35]. The consideration of these properties can inform the design and definition of new policies that consider energy and thermal properties as moving targets and calls for dynamic management and optimization of cloud resources [23]. It is also imperative that the consideration of thermo properties needs to be balanced, traded-off and/or factored along QoS provision; this can be monitored and observed on SLA compliance/violation.

In this work, we consider the thermal characteristics of the host machine focusing on the issue of allocating VMs to hosts in the server farms and assigning workload to the appropriate VMs considering performance parameters [15]. The VMs are sorted based on their resource utilization, memory utilization, disk utilization and network utilization as discussed in *Section 3.2*. The anticipated scheduling policy reduces the energy of the PM, resource utilization with the aid of high-performance distribution strategies [8] [45]. Figure 1 illustrates the basic architecture of resource management in cloud computing environments and describes the interaction of various entities in cloud data center for resource management. The entire incoming load of the server farm is distributed among several VMs for execution [18]. The aggregate workload of the server farm is the finite number of jobs where each job assigned to a few VMs for execution which in turn are hosted by PMs [9].

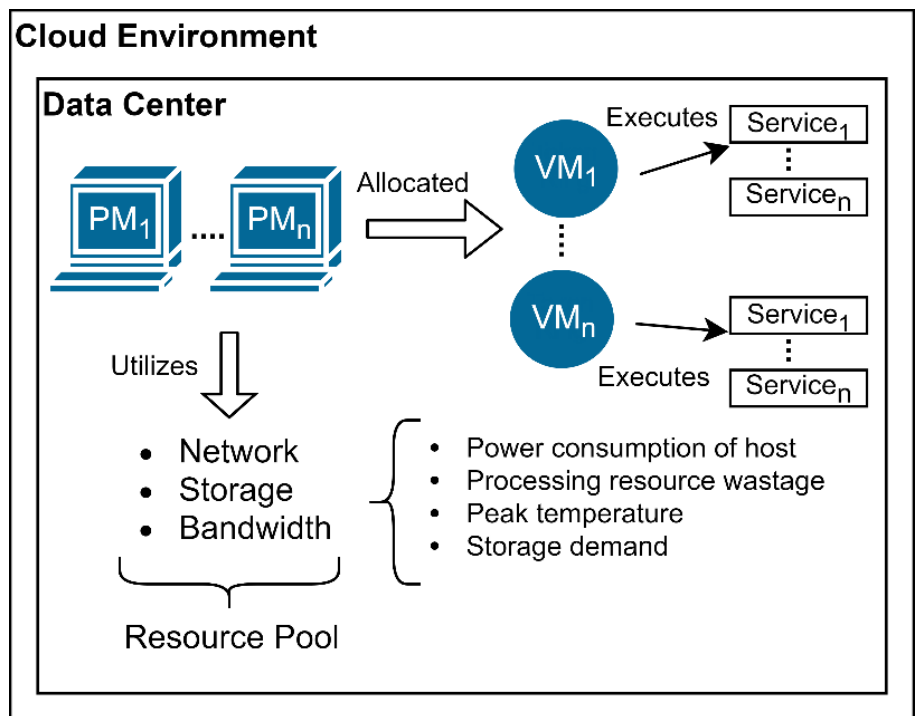

**Fig. 1.** Interaction of various Entities in Cloud Data Center for Resource Management

## 1.1 Motivation and Our Contributions

A well-known cloud simulator, CloudSim toolkit [7] is available, which allows to model and simulates cloud computing environments that resemble real-world infrastructure elements. However, CloudSim toolkit does not include thermal aspects of a data center. Therefore, there is a need to develop a simulation platform as a benchmark to incorporate thermal characteristics which establish a relationship between theory and practice for thermal-aware resource management. The required simulation framework can combine both utilization and thermal models to perform VM allocation to reduce heat generation and hence energy required in computing systems [19] [20] [21] [22] [23]. This more holistic approach allows it to infer more complex patterns of behaviour between resource utilization and heat generation to engender a more efficient approach of resource management which in effect increases the performance of the system [8] [15] [18] [46]. The motivation behind this research work is to propose a framework for the simulation of thermal-aware resource management for cloud computing environment, called ***ThermoSim***, which benefits by combining compute utilization and physical heat characteristics over existing systems. We develop a lightweight deep learning-

                                                                                                    

based temperature predictor using Recurrent Neural Network (RNN) which utilizes resource consumption metrics to accurately predict temperature for cloud hosts. We extend the base classes of CloudSim toolkit to incorporate thermal parameters into it. We evaluated the feasibility and performance of ThermoSim framework and compared it to another baseline simulator.

The ***main contributions*** of this research work are:

1.  A novel framework called ThermoSim is proposed for thermal-aware resource management for cloud computing environments by extending CloudSim toolkit.
2.  In ThermoSim, thermal-aware and utilization based approaches are proposed for scheduling of resources to optimize energy consumption and temperature simultaneously.
3.  A lightweight RNN based deep learning predictor for temperature characterisitcs of cloud hosts is presented for low overhead resource management in ThermoSim
4.  Proposed scheduling approaches equipped with efficient energy and thermal-aware policy for enhanced performance of cloud data centers.
5.  Validated ThermoSim framework against Thas [4] based on different system parameters such as memory, time, cost, energy and prediction accuracy using datasets from Alibaba Cluster and PlanetLab.
6.  ThemoSim analyse the performance of existing energy-aware and thermal-aware scheduling policies based on different QoS parameters such as SLA Violation Rate, energy consumption, number of VM migrations and temperature.
7.  We propose key future research directions in the context of ThermoSim framework.

**Lightweight Testbed/Simulator:** ThermoSim is designed to build an experimental testbed/simulator for conducting practical research in the domain of thermal-aware resource management for cloud computing. Researchers can simulate the thermal behavior of the entire data center using ThermoSim and test or validate their approach for thermal-aware resource management before implementing on real CDC (sensors, servers and GPUs).

### 1.2 Article Organization

The rest of the paper is organized as follows: Section 2 presents the related work. Section 3 presents the ThermoSim framework. Section 4 describes the performance evaluation, validation of ThermoSim framework and experimental results. Section 5 presents conclusions, future research directions and open challenges.

## 2    Related Work

The existing frameworks that allow thermal-aware resource scheduling have significant drawbacks in terms of their ability to extract and predict thermal characteristics in a CDC. We have categorized the related work based on non-SLA and SLA aware resource management techniques for thermal management.

### 2.1 Non-SLA aware Resource Management

Young et al. [51] proposed a Dynamic Thermal Management (DTM) technique, which exploits external computing resources (idle servers) adaptively as well as internal computing resources (free cores of CPU in the server) available in heterogeneous data centers. DTM technique identifies memory intensiveness and usage of VMs if the temperature of a CPU core in a server exceeds a pre-defined thermal threshold and migrates the VMs among CPU cores in the server to maintain the temperature of the server. DTM technique saves energy and improves the performance while satisfying thermal constraints efficiently. DTM technique uses reactive mechanism but not able to predict the temperature variations proactively. Moreover, the impact of temperature variations on SLA has not been identified. Lijun et al. [52] proposed Temperature-Aware Resource Management (TARM) algorithm, which uses Lyapunov Optimization theory to maintain the server temperature without effecting the system reliability. TARM algorithm decreases energy consumed by Computer Room Air Conditioning (CRAC) and servers while imposing server temperature constraints and QoS. Further, a trade-off between system energy consumption and server temperature has been developed, which reported that there is a need of thermal-aware and utilization-based approach to optimize the performance efficiently. TARM algorithm cannot measure the impact of number of VM migrations on system perfromance and SLA.

                    *April 13, 2020*                                        

Lijun et al. [56] proposed a Dynamic Control Algorithm (DCA) without breaking the average temperature constraints and designed a Server Temperature-Constrained Energy Minimization (STCEM) problem. Further, DCA develops linear and quadratic control policies to solve STCEM problem using Lyapunov optimization, similar to TARM algorithm [52]. Further, trade-off between energy and temperature is designed to compare the performance of linear with quadratic control policy and identifies the impact on power usage on system performance during the execution of workloads. İn this approach, the impact of energy consumed by cooling components on overall temerture has not been identified. Zhaohui et al. [58] proposed VM level temperature prediction in Cloud datacenters and measures the impact of CPU temperature on system performance dynamically with/without calibration compared to empirical data. Experimental results show that dynamic CPU temperature modeling with calibration at run time produces more accurate information. In this study, various important performance parameters such as energy, SLA violation rate and their impact on temperature is not discussed. Jean-Marc et al. [60] formulated Mixed İnteger Linear Programming (MILP) for spatio-temporal thermal-aware scheduling while considering dynamics of heat dissipation and production during scheduling of workloads. Further, performance parameters such as energy and execution time are optimized to improve the performance of CDC. İn this research work, the impact of number of VM migrations on system perfromance and SLA is not discussed. Mark et al. [59] proposed an online resource management technique for thermal and energy constrained heterogeneous cloud environment, which uses offline analysis to predict temperature variations at runtime to improve the performance of the system. Earlier temperature prediction helps to execute the workload within their deadline and specified budget. Further, an automatic load balancer is used to balance the load in case of performance degradation while increasing server temperature. This study failed to identify the impact of temperature change on SLA violation rate and energy consumption.

Liu et al [12] proposed a thermal and power-aware model which also considers computing, cooling and task migration energy consumption. But due to their modeling limitations they have not considered I/O energy consumption, network transmission energy consumption. Moreover, they have measured parameters at a very coarse granularity. This work failed to identify the impact of temperature and energy consumption on SLA violation rate and number of VM migrations during execution of workloads. Akbar et al. [13] proposed a game based thermal aware allocation strategy using Cooperative Game Theory and it decreases the imbalance within the CDC by using the concept of cooperative game theory with a Nash-bargaining to assign resources based on thermal profile. A problem in their approach is that they consider a system to be homogeneous and does not incorporate violation rate to categorize tasks and VMs. For heterogeneous environments, their algorithm can lead to contention and poor load balancing. Khaleel [14] described a thermal-aware load balancing strategy by calculating the shortest distance to cloud resources deployed at different geographical locations and conserving more bandwidth cost. Further, it has been suggested that running servers at different locations can reduce temperature at particular location and improves the health of server. Proposed strategy improves the resource utilization without considering the other type of utilization such as disk, network and memory. This would not be feasible in virtual containers-based cloud services and mobile service providers.

## 2.2 SLA aware Resource Management

Ying et al. [57] proposed thermal-aware VM migration manager to identify the working condition of server based on resource utilization and temperature and recognizes the impact of CPU overheating (caused by chassis fan damage) on system performance. Further, it is shown that the migration of VMs from overloaded server to underloaded server balances the load, reduces the damages to overloaded servers and improves its health and system performance. Proposed technique decreases the number of VM failures and improves the system ability which further reduces SLA violation rate, but this technique has not been identified the impact of temperature variation on SLA violation rate. Mhedheb et al. [4] proposed a Thermal aware scheduler (Thas) on the CloudSim toolkit, which implements an interface between the hypervisor and the virtual machine in CDC. They have replaced the CloudSim's VMScheduler class which results in their scheduling behavior to be highly data-dependent of the existing class inputs. However, Thas lacks thermal characteristic studies by which they are not able to find the best host location at the time of VM migration. Moreover, their model is restricted in terms of analyzing the CPU and memory loads and the number of VM migrations which significantly impacts the performance of the system in terms of its latency and other QoS parameters. Xiang et al. [8] proposed a virtual machine scheduling technique for Cloud Datacenters which reduces total energy consumption while holistically managing the components of CDC such as CPU, memory, storage, network and cooling. This research work proposed a thermal and energy-aware model to analyze the temperature variation and distribution in the CDC during execution of workloads. Further, a holistic resource management approach is proposed to reduce the energy consumption of CDC and maintains the temperature of CDC below than critical temperature. The architecture comprises of three sub-components such as workload manager, scheduling manager and cooling manager. Workload manager manages the workloads submitted by user and process for scheduling based on their requirements. Scheduling manager schedules the resources for execution of workloads while maximizing the performance of CDC and minimizes the consumption of energy. Cooling manager maintains the temperature of CDC and save cooling energy by performing VM placement and dynamic migration in an efficient manner. Proposed model updates computing capacity dynamically to improve cooling efficiency and maintains the CPU temperature less than threshold value while adjusting cooling energy to the lowest level to save energy and reduce SLA violation rate. It uses reactive mechanism to maintain temperture but not able to predict the temperature variations proactively. Moreover, the impact of temperature variations on SLA has not been identified.

Rodero et al. [10] also provided a strategy to allocate VMs using their temperature characteristics. They propose a reactive technique as an alternative to VM migration and DVFS which reduces the activity of one or more VMs by pinning them to specific Cloud Management Portals (CMPs). They profile applications to decide which VM to pin with which CPU. This in effect has significant overhead in terms of CPU, Memory and Time and not able to exploit thermal data completely as it significantly depends on the OS's default dynamic CPU power management which is not aware of the characteristics of other physical machines in the network. Kumar et al. [11] presented a suite of heuristics for energy efficiency consolidation and a hybrid scheduling algorithm to maintain the temperature of the CDC and reduces the energy consumption of different servers within the CDC. They propose StaticPPMMax (performance to power metrics using server's peak power consumption) and compute metrics for VM allocation, but do not consider server process sleep state and transition power consumption which makes their approach weak and not scalable for complex workload models. Ilager et al. [15] proposed an Energy and Thermal-Aware Scheduling (ETAS) algorithm that dynamically consolidates VMs to minimize the overall energy consumption while proactively preventing hotspots. They have extended a class of the CloudSim toolkit which does not allow them to validate their results on fair grounds. Moreover, their algorithm assumes static cooling environment, which may not be versatile to different cooling settings.

## 2.3 Critical Analysis

Table 1 shows the comparison of ThermoSim with existing frameworks. All the above research works have presented thermal-aware scheduling frameworks in cloud computing without considering the thermal-aware and utilization-based resource management simultaneously in a single framework, but it is very important to study the behaviour of both resource management approaches together to optimize the different QoS parameters in a controlled and holistic manner. None of the existing works validated against prediction accuracy, memory and time and only two frameworks [14] [15] considered time for validation. As per literature, there is a no existing framework which considers all the four QoS parameters (SLA Violation Rate, energy consumption, number of VM migrations and temperature) in a single framework. Due to this, the current thermal-aware resource management become inefficient to respond in these situations. Thermal aware techniques provide benefits in some cases, however fail in some other cases for which energy aware approaches are required [18] [31] [32] [35]. Thus, ThermoSim uses an integrated approach leveraging both techniques for optimum results.

**Table 1.** Comparison of ThermoSim with exiting frameworks

| Framework | Performance Parameters | | | | Validation | | | | | Deep Learning based Temperature Prediction | Resource Management | |
|---|---|---|---|---|---|---|---|---|---|---|---|---|
| | SLA Violation Rate | Energy Consumption | Number of VM Migrations | Temperature | Prediction Accuracy | Memory | Time | Cost | Energy | | Thermal-aware Approach | Utilization based Approach |
| Non-SLA aware Resource Management | | | | | | | | | | | | |
| Liu et al. [12] | | ✓ | | ✓ | | | | | ✓ | | ✓ | |
| Akbar et al. [13] | | ✓ | ✓ | ✓ | | | | | ✓ | | ✓ | |
| Khaleel [14] | | ✓ | ✓ | | | | ✓ | | | | ✓ | |
| Young et al. [51] | | ✓ | | ✓ | | | | | | | ✓ | |
| Lijun et al. [52] | | ✓ | | ✓ | | | | | | | ✓ | |
| Lijun et al. [56] | | ✓ | | ✓ | | | | | | | ✓ | |
| Zhaohui et al. [58] | | | | ✓ | | | | | | | ✓ | |
| Jean-Marc et al. [60] | | ✓ | | | | | ✓ | | | | ✓ | |
| Mark et al. [59] | | | | ✓ | | | | ✓ | | | ✓ | |
| SLA aware Resource Management | | | | | | | | | | | | |
| Y. Mhedheb et al. [4] | ✓ | ✓ | | ✓ | | | | | ✓ | | ✓ | |
| Rodero et al. [10] | ✓ | ✓ | ✓ | ✓ | | | | | ✓ | | ✓ | |
| Kumar et al. [11] | ✓ | ✓ | | ✓ | | | | | ✓ | | ✓ | |
| Ilager et. al [15] | ✓ | ✓ | ✓ | ✓ | | | ✓ | | ✓ | | ✓ | |
| Ying et al. [57] | ✓ | ✓ | ✓ | | | | | | | | ✓ | |
| ThermoSim | ✓ | ✓ | ✓ | ✓ | ✓ | ✓ | ✓ | ✓ | ✓ | ✓ | ✓ | ✓ |

Therefore, there is a need to develop a simulation platform as a benchmark: 1) to establish a relationship between theory and practice for thermal-aware resource management, 2) to incorporate thermal characteristics, 3) to schedule resources using thermal-aware and utilization-based resource management simultaneously, 4) to validate the thermal-aware resource management framework against system parameters such as memory, time, cost, energy and prediction accuracy and 5) to test the performance of thermal and energy-aware scheduling policies based on all the four QoS parameters such as SLA Violation Rate, energy consumption, number of VM migrations and temperature. Our proposed ThermoSim framework addresses the challenges of existing frameworks in this research work.

## 3 ThermoSim Framework

This section presents the detailed description of TheromSim framework. Figure 2 presents the architecture of the ThermoSim framework, which is based on two different models: 1) Utilization Model 2) Thermal Model. Energy model is an integral part of Utilization model. In the utilization model, cloud workloads are assigned to VMs based on their different types of utilization (resource, network, memory and disk) and energy consumption, while the thermal model considers the thermal characteristics of the host machine and accordingly VMs are scheduled on PMs.

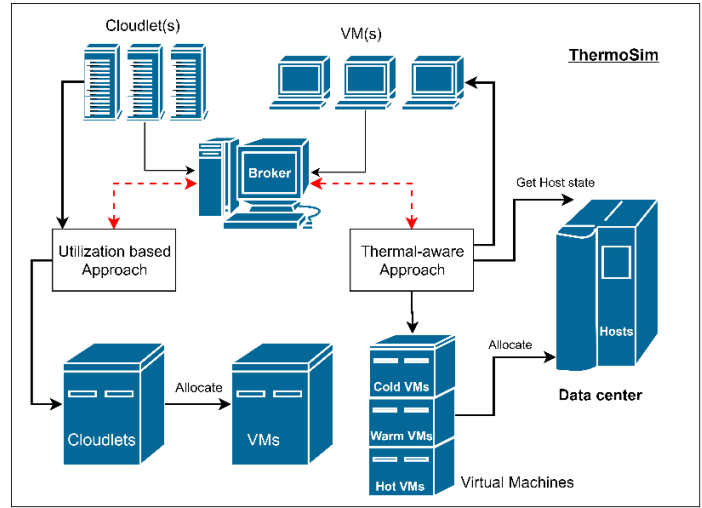

**Fig. 2.** Architecture of ThermoSim Framework

### 3.1 Energy Model

The larger part of the data center energy consumption is contributed by the computing and cooling systems [3] [7] [18] [20] [21] [25] [31] [32] [33] [34].

$$E_{Total} = E_{Computing} + E_{Cooling} \qquad (1)$$

**3.1.1 Computing:** The computing system consists of hosts and its energy consumption can be defined as follows:

$$E_{Computing} = E_{Processor} + E_{Storage} + E_{Memory} + E_{Network} + E_{Extra} \qquad (2)$$

$E_{Processor}$ represents the processor's energy consumption, which is calculated using [Eq. 3]:

$$E_{Processor} = \sum_{r=1}^{r=cores} (E_{dynamic} + E_{SC} + E_{Leakage} + E_{idle}) \qquad (3)$$

where $E_{dynamic}$ represents dynamic energy consumption, $E_{SC}$ represents short-circuit energy consumption, $E_{Leakage}$ represents power loss due to transistor leakage current and $E_{idle}$ represents the energy consumption when processor component is idle. Dynamic energy consumption ($E_{dynamic}$) is calculated using [Eq. 4], which is an average of energy consumption calculated using linear and non-linear model.

$$E_{dynamic} = \frac{E_{dynamic}^{linear} + E_{dynamic}^{non-linear}(U_j)}{2} \qquad (4)$$

$E_{dynamic}^{linear}$ is the dynamic energy using linear model [8] and it is calculated using [Eq. 5]:

$$E_{dynamic}^{linear} = CV^2 f \qquad (5)$$

where $C$ is capacitance, $f$ is frequency, and $V$ is voltage. $E_{dynamic}^{non-linear}$ is the dynamic energy using non-linear model [43], resource utilization has a non-linear relationship with energy consumption and it is calculated using [Eq. 6]:

$$E_{dynamic}^{non-linear}(U_j) = \mu_1.U_j + \mu_2.U_j^2 \qquad (6)$$

Where $\mu_1$ and $\mu_2$ are nonlinear model parameters and $U_j$ is CPU utilization of host $h_j$.

$E_{Storage}$ represents the energy consumption of storage device, which performs data read and write operations and it is calculated using [Eq. 7]:

$$E_{Storage} = E_{ReadOperation} + E_{WriteOperation} + E_{idle} \qquad (7)$$

where $E_{idle}$ represents the energy consumption when storage component is idle.

$E_{Memory}$ represents the energy consumption of the main memory (RAM/DRAM) and cache memory (SRAM), which is calculated using [Eq. 8]:

$$E_{Memory} = E_{SRAM} + E_{DRAM} \qquad (8)$$

$E_{Network}$ represents the energy consumption of networking equipment such as routers, switches and gateways, LAN cards, which is calculated using [Eq. 9]:

$$E_{Network} = E_{Router} + E_{Gateway} + E_{LANcard} + E_{Switch} \qquad (9)$$

$E_{Extra}$ represents the energy consumption of other parts, including the current conversion loss and others, which is calculated using [Eq. 10]:

$$E_{Extra} = E_{Motherboard} + \sum_{f=0}^{F} E_{connector,f} \qquad (10)$$

where $E_{Motherboard}$ is energy consumed by motherboard (s) and $\sum_{f=0}^{F} E_{connector,f}$ is energy consumed by a connector (port) running at the frequency $f$.

**3.1.2 *Cooling:*** The energy model for computing is developed based on energy consumption of different cooling components to maintain the temperature of CDC. $E_{Cooling}$ represents the energy is consumed by cooling devices (compressors, Air Conditioners (AC) and fans) to maintain the temperature of cloud datacenter, which is calculated using [Eq. 11]:

$$E_{Cooling} = E_{AC} + E_{Compressor} + E_{Fan} \qquad (11)$$

## 3.2 Utilization Model

The jobs originate by cloud consumers as demanded services known as "cloud workloads" [26]. These workloads are submitted to the workload queue of the cloud system. The workload is sorted in ascending order of VMs based on resource utilization and placed in the queue to assign workload to virtual machines. *Resource Utilization* is a ratio of an execution time of a workload executed by a particular resource to the total uptime of that resource [18]. The total uptime of resource is the amount of time available with a cloud resource set for execution of workloads. We have designed the following formula to calculate resource utilization ($R_{Utilization}$) [Eq. 12].

$$R_{Utilization} = \sum_{i=1}^{n} \left( \frac{execution\ time\ of\ a\ workload\ executed\ on\ i^{th}\ resource}{total\ uptime\ of\ i^{th}\ resource} \right) \qquad (12)$$

Where $n$ is the number of resources. These VMs are also sorted in terms of their network utilization, memory utilization and disk utilization in descending order i.e. opposite order in which workloads are sorted and placed in the queue. The formula for calculating memory utilization ($M_{Utilization}$) in percentage [18] is as follows [Eq. 13]:

$$M_{Utilization} = \frac{\text{Total Physical Memory} - (\text{Memory Free} + \text{Memory Buffers} + \text{Cache Memory})}{\text{Total Physical Memory}} \times 100 \qquad (13)$$

The formula for calculating disk utilization ($D_{Utilization}$) in percentage is as follows [Eq. 14]:

$$D_{Utilization} = \frac{\text{Total Used}}{\text{Total HD size}} \times 100 \qquad (14)$$

$$D_{Utilization} = \frac{\text{Storage Allocation Units} \times \text{Storage Used}}{\text{StorageAllocationUnits} \times \text{Storage Size}} \times 100 \qquad (15)$$

The formula for calculating network utilization ($N_{Utilization}$) in percentage [18] is as follows [Eq. 16]:

$$N_{Utilization} = \frac{\text{data bits}}{\text{bandwidth} \times \text{interval}} \times 100 \qquad (16)$$

The workloads present in the sub task queues are submitted to the data center broker. Workload classification and assignment are shown in Figure 3. Thus, cloud workloads are mapped to the VMs based on energy-efficient resource management policy using Cuckoo Optimization based scheduling technique [18]. [Algorithm 1] presents the utilization-based approach which i) sorts the cloud workloads and ii) maps workloads to VMs. Based on the current utilization of VMs, the scheduler allocates the tasks accordingly. To do this, workloads are organized in increasing order of utilization, and VMs in decreasing order with an aim to allocate tasks which lower the utilization i.e. light-weight tasks are executed on VM with high utilization, and vice-versa.

| Algorithm 1: Utilization based Approach for Resource Management |
|---|
| 1. **Input:** Number of workloads (T) and number of available resources (V) |
| 2. **Output:** Mapping of each workload to the resource |
| 3. **Function** *UtilizationSort (list, vm, decreasing)* |
| 4.   **if** (vm == True) **then** |
| 5.     Sort list based on increasing order of $E_{Total}$ |
| 6.     **return** (*UtilizationSort* (list, false)) |
| 7.   **if** (decreasing == True) **then** |
| 8.     Sort list based on decreasing order of $R_{Utilization}$ |
| 9.   **else** |
| 10.     Sort list based on increasing order of $R_{Utilization}$ |
| 11.   Break ties using $M_{Utilization}$ |
| 12.   Break ties using $D_{Utilization}$ |
| 13.   Break ties using $N_{Utilization}$ |
| 14.   **return** (list) |
| 15. **Start** |
| 16.   Initialize all host list (Number of PMs) |
| 17.   Initialize all resource list (Number of VMs) |
| 18.   Initialize all workload list (Number of tasks) |
| 19.   T' = *UtilizationSort* (T, false, false) // sort task with increasing utilization (estimating variation in utilization while executing task *t*) |
| 20.   V' = *UtilizationSort* (T, true, true)  // sort VM with decreasing utilization (function to estimate utilization for VM) |
| 21.   **for** task *t* in T: |
| 22.     **for** vm *v* in V: |
| 23.       **if** *t* is suitable for *v* in V **then** |
| 24.         Schedule the task *t* on VM *v* |
| 25. **End** |

The main idea of Algorithm 1 is described below:

1. Initialize the hosts (PMs) as available at time stamp to *0* i.e. host is available at beginning of scheduling.
2. Sort workloads and VMs based on utilization.
3. Map workloads to VMs.

Algorithm 1 sorts tasks with increasing utilization requirements and sorts VMs with decreasing utilization requirements. This allows high resource requirement tasks to be allocated on VMs which have low load. This greedy algorithm allows efficient scheduling of tasks to VMs based on their requirements. The sorting algorithm can be implemented using merge sort [O(n log n)], and the allocation of tasks we iterate over all VMs for every task which makes it [O(n$^2$)] in worst case. Here $n$ is maximum of the number of tasks or VMs.

### 3.3 Thermal Model

The idea is to design a scheduling policy for VMs based on the CPUs temperature characteristics. Thus, a thermal model is needed that describes these changes of this parameter when workloads are running on virtual machines. Thermal-aware scheduling considers current temperature and maximum working temperature, that is, the threshold temperature of every machine [15] [19] [20] [21] [22] [27], before making scheduling decisions. Let the maximum threshold temperature of a server machine be $T_{over}$ and let current temperature of a server machine be $T_{cu}$. $T_{over}$ is the temperature beyond which a machine is overheated. The heuristic chosen for VM scheduling is the difference between threshold and present temperature, as formulated in [Eq. 17]. $\Delta T_{vi}$ is the temperature variation of the host.

$$\Delta T_{vi} = T_{over} - T_{cu} \qquad (17)$$

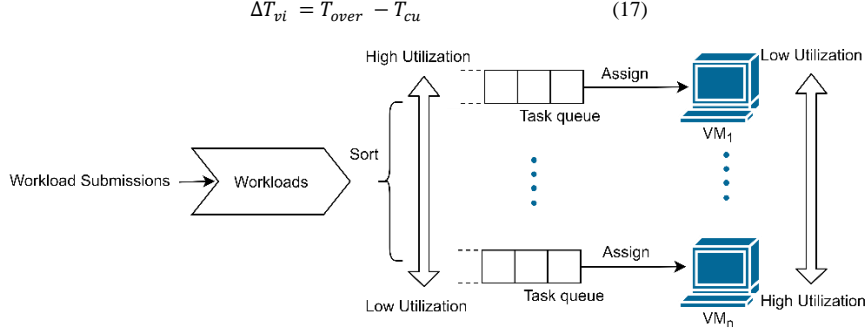

**Fig. 3.** Workload Classification and Assignment

We used Computer Room Air Conditioning (CRAC) model and RC (where R and C are thermal resistance (k/w) and heat capacity (j/k) of the host respectively) thermal model [17] [18] [19] [20] [21] [22] [31] to design temperature model for calculation of current temperature of CPU ($T_{cu}$). The following formula is used to calculate the current temperature of CPU [Eq. 18].

$$T_{cu} = PR + Temp_{inlet} + T_{initial} \times e^{-RC} \qquad (18)$$

where CRAC model is used to calculate inlet temperature ($Temp_{inlet}$) and RC model is used to calculate CPU temperature ($Temp_{CPU}$) and $P$ is the dynamic power of host. $T_{initial}$ is the initial temperature of the CPU. The relationship between a power (energy efficiency) and temperature model (heating model) [23] is described in [Eq. 18]. VMs are separated into various classes as per their temperature attributes, which are then distributed to the hosts based on their temperatures. The VM movement component is directed to guarantee the unwavering quality of the model when a host achieves threshold temperature.

### 3.4 Deep Leaning based Temperature Prediction Module

In many CDCs, it is difficult to access the thermal characteristics of the hosts [45]. Mostly, the temperature sensors are either too expensive or are unavailable to give accurate temperature information [40] [41]. Having noisy temperature data can significantly affect the performance of any thermal-aware simulator including ThermoSim. The thermal model in Section 3.3 requires various metrics like thermal resistance (R), capacitance (C), $T_{initial}$, which may not be available for many CDCs. Moreover, the CRAC and RC models are resource intensive. Thus, there is a requirement of a prediction module that can predict thermal characteristics by observing simpler metrics like memory, CPU, I/O utilization and fan speeds. ThermoSim, thus, can also use a temperature prediction module that uses Recurrent Neural Network (RNN) [42] to predict CPU temperatures for different PMs. This is not only useful for CDC where temperature sensors are unavailable or expensive but also when the sensor data is noisy.

To predict the CPU temperature, our deep learning model uses RNN with 4 Gated Recurrent Units (GRU) layers as shown in Figure 4. The input of the network is a matrix with various features of all PMs. These features include fan speeds, and resource utilization metrics. The output of the network is a vector of temperatures for all PMs. A Recurrent Neural Network (RNN) is a class of artificial neural networks where connections between nodes form a directed graph along a temporal sequence. This allows it to exhibit temporal dynamic behavior. Unlike feedforward neural networks, RNNs can use their internal state (memory) to process sequences of inputs. This makes them applicable to tasks such as unsegmented, connected handwriting recognition or speech recognition. However, analyzing temperature characteristics also requires the exploitation of temporal workload and processing patterns, which can be done best with RNNs. However, RNNs face the problem of vanishing gradients which makes the network updates very slow, for which the research community has introduce GRUs. GRUs are a gating mechanism in recurrent neural networks, introduced in 2014 by Kyunghyun Cho et al. [39]. To solve the vanishing gradient problem of a standard RNN, GRU uses, so-called, update gate and reset gate. These are two vectors which decide what information should be passed to the output. The special thing about them is that they can be trained to keep information from long ago, without washing it through time or remove information, which is irrelevant to the prediction.

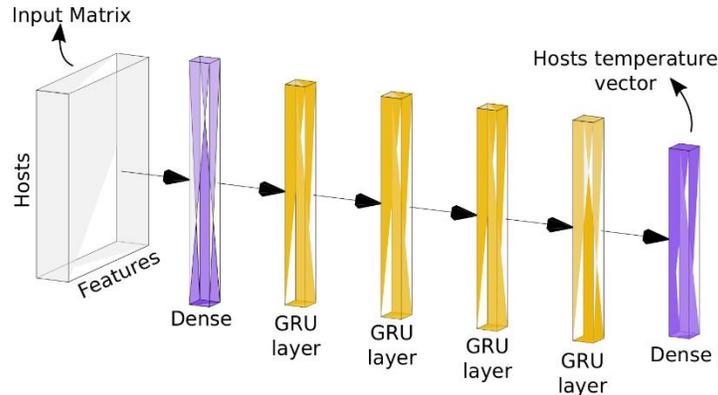

**Fig. 4:** Architecture of RNN to predict temperature of PMs

To train the neural network, we used traces[1] from Alibaba CDC, a large-scale ecommerce provider [47] [48] [49] [50]. Dataset extracted from traces on a cluster of 70 servers with 64-bit Ubuntu 18.04 Operating System, every server is equipped with the Intel® Core™ i7 9700k processor (No. of Cores 8, No. of Threads 8, Processor Base Frequency 3.60 GHz, Max Turbo Frequency 4.90 GHz, Max Cache 12 MB Intel® Smart Cache, Bus Speed 8 GT/s and TDP 95 W), 16 GB of RAM, 512 GB SSD storage and Graphics card GTX 2060. Hadoop MapReduce is installed on all the servers for processing to execute word count application. Every server hosts at least 15 virtual machines or nodes. A sample from the dataset is shown in Table 2. However, only utilization of various elements of the server is not sufficient for accurate prediction of temperature. Previous works show that Fan speeds also affect the server temperature [16] [21]. For our experiments, we have taken configuration with 5 fans, so we calculate Fan speed from the different utilization metrics in the dataset. Fan speed is calculated using [Eq. 19], which is multiplication of average utilization ($Avg_{\text{utilization}}$) and CPU temperature ($Temp$) [16] [21] [34] [53].

$$RPM_{Fan} = \alpha(Avg_{\text{utilization}} \times Temp) \qquad (19)$$

where $\alpha$ = 1.5 RPM/Degree Celsius

For 5 fans, we keep fan speed $RPM_{Fan}$, a random value between $[RPM_{Fan} - \frac{1}{2}\Delta RPM_{Fan}, RPM_{Fan} + \frac{1}{2}\Delta RPM_{Fan}]$, where $\Delta RPM_{Fan}$ is defined using [Eq. 20].

$$\Delta RPM_{Fan} = \alpha(\Delta Avg_{\text{utilization}} \times Temp + Avg_{\text{utilization}} \times \Delta Temp) \qquad (20)$$

As per the dataset, the precision of $Avg_{\text{utilization}}$ within 1% and precision of Temp is within 1 Degree Celsius (°C). Using 1100 training and 100 test datapoints, our network is able to reach accuracy of 96.78%.

**Table 2.** Sample data extracted from Alibaba CDC

| Server ID and Time | | Fan (RPM) | | | | | Utilization (%) | | | | Temp (°C) |
| --- | --- | --- | --- | --- | --- | --- | --- | --- | --- | --- | --- |
| Id | Time Stamp | F1 | F2 | F3 | F4 | F5 | System | Memory | CPU | I/O | CPU |
| N1 | 8:29:49 | 4214 | 4289 | 4230 | 4264 | 4263 | 58 | 62 | 63 | 72 | 44 |
| N2 | 6:09:23 | 3979 | 4046 | 4085 | 4060 | 4033 | 67 | 72 | 35 | 84 | 42 |

[1] Alibaba Cluster - https://github.com/alibaba/clusterdata

| N3 | 7:20:22 | 4389 | 4403 | 4311 | 4286 | 4386 | 43 | 76 | 61 | 40 | 53 |
|----|---------|------|------|------|------|------|----|----|----|----|----|
| N4 | 6:53:08 | 5013 | 4928 | 5001 | 4981 | 5099 | 65 | 67 | 50 | 62 | 55 |
| N5 | 6:30:09 | 3552 | 3635 | 3601 | 3591 | 3499 | 67 | 82 | 30 | 38 | 44 |
| N6 | 6:01:46 | 4012 | 3970 | 3939 | 3891 | 3919 | 66 | 69 | 64 | 50 | 42 |

Using this deep leaning-based prediction module, ThermoSim can also classify VMs and PMs based on thermal-aware technique even for CDCs where temperature sensors are unavailable.

### 3.5 Virtual Machine Classification and Scheduling

Figure 5 presents the process of virtual machine classification and allocation. Virtual machines are categorized into three classes in accordance with their thermal features: hot, warm, and cold. A 'cold' VM symbolizes that a VM may decrease the temperature of a PM if its temperature is greater than $\theta_{vl}$. A 'hot' VM states that a VM may raise the temperature of a host if its temperature is greater than $\theta_{vh}$. A 'warm' VM signifies the temperature of a PM will remain stable.

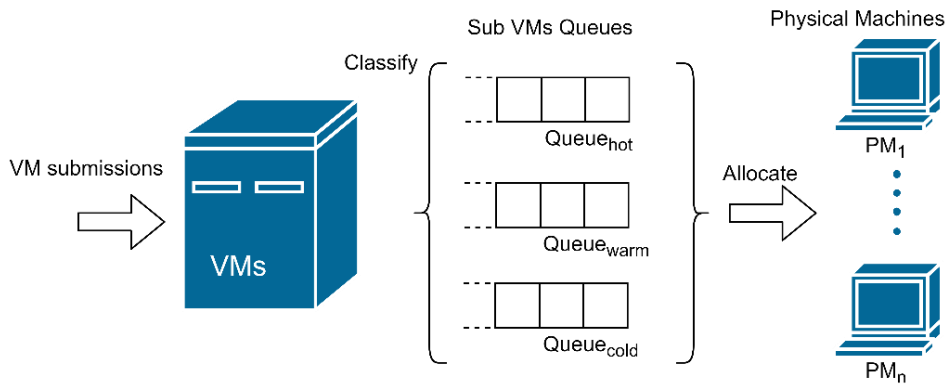

**Fig. 5:** Process of Virtual Machine Classification and Allocation

The thermal scheduler allocates VMs to the PM whose temperature is farthest away from its maximum threshold temperature. The scheduler also manages a waiting queue. It is the queue through which demand for a new VM is fulfilled. Each new request for VM is added at the end of waiting queue [16].

| Algorithm 2: Thermal-aware Approach for Resource Management |
|---|
| 1.   **Input:** Number of VMs and number of available PMs |
| 2.   **Output:** Scheduling of VMs to the PMs |
| 3.   **Start** |
| 4.   Initialize all host list (Number of PMs) |
| 5.   Initialize all VMs list (Number of Resources (VMs) - vmlist) |
| 6.   **do** |
| 7.   { |
| 8.       dequeue VM from queue |
| 9.       calculate $\Delta T_{vi}$ |
| 10.    **if** ( $\Delta T_{vi} > \theta_{vh}$ ) **then** |
| 11.        enqueue the VM to $Q_{Hot}$ |
| 12.    **else if** ($\Delta T_{vi} < \theta_{vl}$) **do** |
| 13.        enqueue the VM to $Q_{cold}$ |
| 14.    **else if** ( $\theta_{vh} > \Delta T_{vi} > \theta_{vl}$) **do** |
| 15.        enqueue the VM to $Q_{warm}$ |
| 16.    } |
| 17.  **while** !vmlist.empty() |
| 18.  **if** ($T_{over} > \theta_{th}$ and ! $Q_{cold}$.empty()) **then** |
| 19.      dequeue($Q_{cold}$); |
| 20.  **else if** (!$Q_{warm}$.empty()) **then** |
| 21.      dequeue($Q_{warm}$); |
| 22.  **else** |
| 23.      dequeue($Q_{Hot}$); |
| 24.  **if** ($T_{over} < \theta_{th}$ & ! $Q_{Hot}$.empty()) **then** |
| 25.      dequeue($Q_{Hot}$); |
| 26.  **else if** (!$Q_{warm}$.empty()) **then** |
| 27.      dequeue($Q_{warm}$); |
| 28.  **else** dequeue($Q_{cold}$); |
| 29.  allocate VM to PM |
| 30.  **End** |

The scheduler will remove that VM request from the waiting queue as per further requirement; then further scheduler queues the VMs into sub-queues based on the temperature variation of the host due to the VM i.e., value of $\Delta T_{vi}$. If the temperature variation of host is greater than the high temperature threshold, then the VM is added to the hot queue; if the temperature variation of host is less than the low temperature threshold, then it is added to the cold queue, otherwise it is added to the warm queue [5] [16]. For VM allocation, comparison is done: if the current temperature of a host is more than the high temperature threshold of a host, a cold VM from the cold queue will be allocated [6]. Subsequently, the host temperature lowers to the ordinary state and a VM will be allocated on the host to execute. When the temperature of a host is greater than $\theta_{ch}$ and VM from the cold queue is allocated to the host, the temperature then drops to below $\theta_{ch}$. When the temperature of a host is less than $\theta_{cb}$, a hot VM will be selected to execute. For cases that the temperature of a host is between $\theta_{cl}$ and $\theta_{ch}$, the VM from warm queue will be executed. [Algorithm 2] presents the thermal-aware approach, which contains two phases: 1) sorting and 2) resource scheduling.

The thermal-aware algorithm sorts the VMs based on their execution effect on PMs. Algorithm 2 executes hot VMs on cold nodes and cold VMs on hot nodes and tries to reduce energy consumption. The symbols used in the [Algorithm 2] are defined in Table 3.

Table 3: Symbols and its description

| Symbol | Description |
|---|---|
| $\Delta T_{vi}$ | Temperature variation of host due to execution of current VM |
| $T_{over}$ | Temperature of overheated host |
| $T_{normal}$ | Temperature of normal host |
| $T_{danger}$ | Temperature of overheating host |
| $\theta_{vh}$ | The low temperature threshold of $\Delta T_{vi}$ |
| $\theta_{vl}$ | The high temperature threshold of $\Delta T_{vi}$ |
| $\theta_{cl}$ | Low temperature threshold of host // normal host temperature |
| $\theta_{ch}$ | High temperature threshold of host //overheating host temperature |

For calculating the value of $\theta_{vl}$ and $\theta_{vh}$ different types of VMs according to temperature variation of the host machine;

$$\theta_{vh} = T_{over} - T_{danger} \qquad (21)$$

$$\theta_{vl} = \frac{1}{2} T_{normal} - T_{danger} \qquad (22)$$

Several papers from the literature [2] [5] [6] [9] [15] [23] reported a different range of values for temperature:

- $T_{over}$ = 79°C. // overheated host temperature
- $T_{danger}$ = 70°C. //overheating host temperature
- $T_{normal}$ = 29°C. // normal host temperature

The main idea of Algorithm 2 is described below:

1. Initialize the all hosts (PMs) as available at time stamp to *0* i.e. host is available at the beginning of resource scheduling.
2. Calculate the temperature variation, before and after allocating the VM until queue is not empty.
3. Allocate VMs to sub-queues according to the temperature variation value of the host.
4. Dequeues VMs from their sub-queues according to the host state and allocates VM to PM.
5. Execute the allocated VM on PM.

Algorithm 2 maintains 3 sub-queues for VMs, classified into hot, warm and cold. The VMs are allocated to these queues based on their temperature models. Based on the queues and temperatures of hosts, we allocate hot VMs to cold hosts and cold ones to hot hosts. If hot or cold VM queues are empty, we next consider warm queues. This way the temperature remains balanced. The overall complexity of the algorithm is O(n) where *n* is maximum of number of hosts and VMs.

### 3.6 Integration of All Models

The utilization model is used in integration with the energy model. It provides quantified values for resource utilization for different computational elements like CPU, memory and disk. The energy model is used to calculate the energy consumption for different components classified broadly into computing and cooling. Both energy and utilization models are used for resource management of allocating tasks to VMs in ThermoSim as described in Algorithm 1. Moreover, the thermal model using temperatures calculated by either [Eq. 18] or the RNN based prediction module is used for thermal-aware approach for resource management of allocating VMs to PMs, as shown in Algorithm 2. We show use cases of both algorithms in performance evaluation, where we compare energy-based and thermal-based resource allocation policies as described in Section 4.

## 4    Performance Evaluation

This section describes the performance evaluation, configuration details, type of workload and experimental results for validation of ThermoSim. The CloudSim toolkit [7] has been extended by developing a new ThermoSim framework to incorporate thermal parameters. Figure 6 shows the class diagram where the ThermoSim framework is extending different classes (VMAllocationPolicy, VMScheduler, DataCenterCharacteristics and CloudletScheduler) from the CloudSim toolkit and developed four different classes (ThermoCloud, UtilizationBased, ThermalAware and *TemperaturePredictor*). *ThermoCloud* is main class, which interacts with classes of CloudSim toolkit and controls *UtilizationBased* class, *ThermalAware* class and *TemperaturePredictor* class. Figure 7 shows the sequence diagram, which describes the interaction among various classes of ThermoSim framework during execution of workloads.

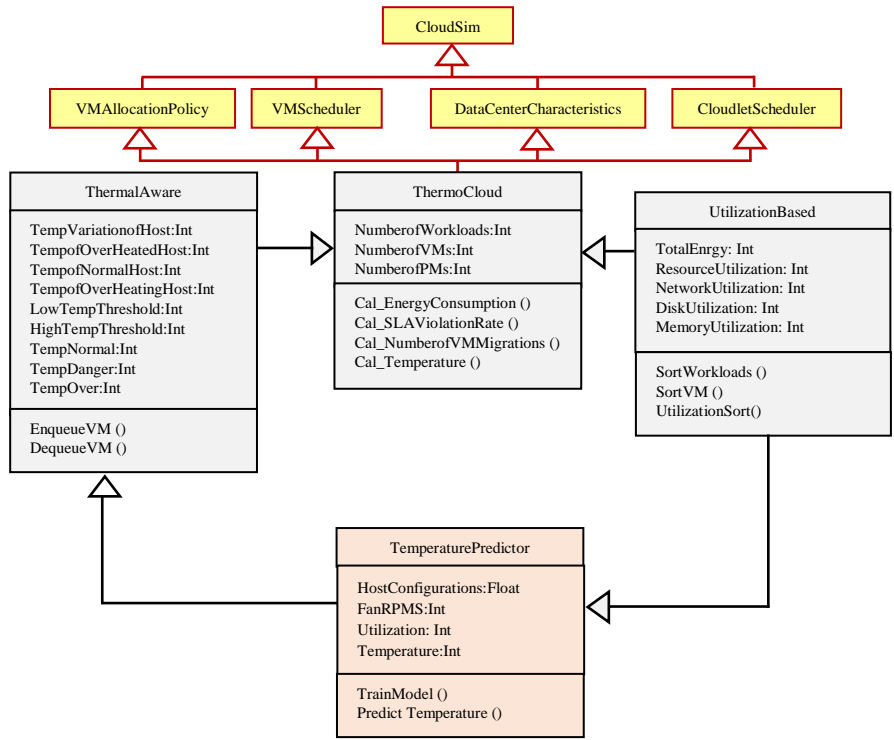

**Fig. 6.** ThermoSim framework is extending different classes from the CloudSim Toolkit

### 4.1 Configuration Settings

We have conducted experiments on a machine with an Intel® Core™ i7-7820HQ Processor (8M Cache, 3.90 GHz), 16 GB RAM and 2 TB of HDD running on 64-bit Windows OS. Our CDC comprises of 4 PMs with configuration (Cores = 4, CPU MIPS = 2000, RAM = 8 GB, Bandwidth = 1 Gbit/s) and 12 VMs with configuration (Core = 1, CPU MIPS = 500, RAM = 1 GB and Bandwidth = 100 Mbit/s). We have run the experiments for 10 times and the average results are reported. We have run the simulation for periods of 48 hours and executed the scheduling algorithm after every 5-minute interval for dynamic consolidation of VMs. Virtual nodes are further divided into instances called Execution Components (ECs). Every EC contains their own cost of execution and it is measured with unit (C$/EC time unit (Sec)). EC measures cost per time unit in Cloud dollars (C$).

For the execution of workloads in our experiments, we have chosen varied computational settings on top of heterogeneous resources. The variety comes in the number of cores at the CPU-level, the page levels of the main memory, switches at the network level and disk space at the storage level [7] [28] [29] [18]. Cores is the number of Processing Element's (PE) required by the Cloudlet. Table 4 shows the simulation parameters utilized in the various experiments undertaken by this research work, also as identified from the existing empirical studies and literature such as utilization model [20] [30], energy model (computing [3] [18] [32] [33] [37] [38] and cooling [20] [21] [31] [34]) and thermal-aware scheduling [21] [22] [31]. Experimental setup incorporated CloudSim to produce and retrieve simulation results.

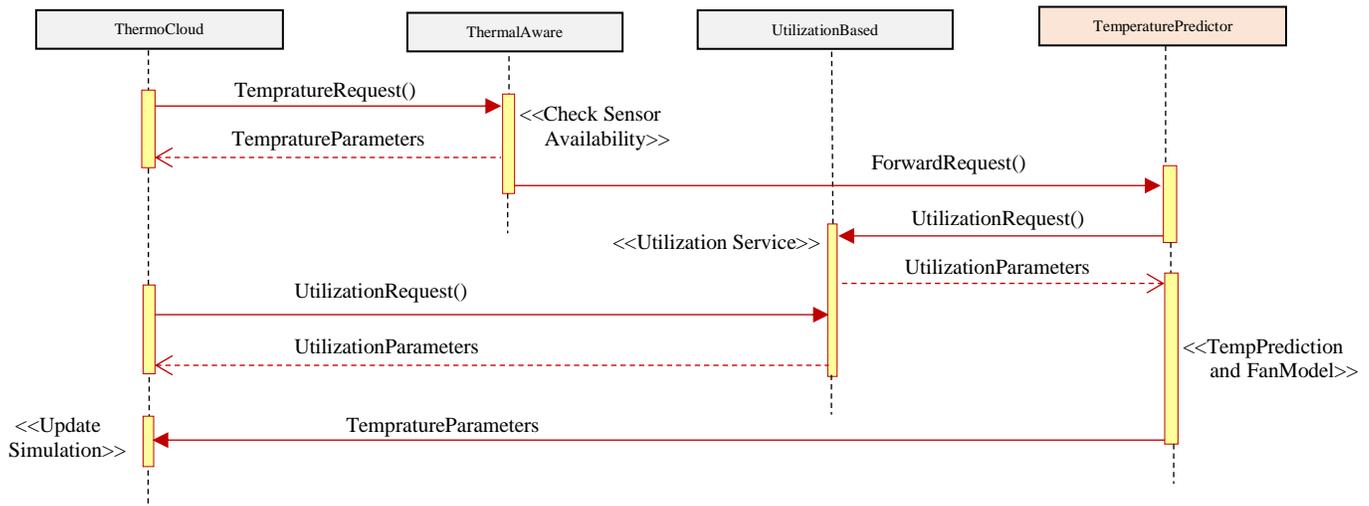

**Fig. 7.** Sequence diagram describes interaction among various classes of ThermoSim framework

### 4.2 Workload

The ThermoSim framework uses PlanetLab[2] dataset [44] and it is considered as a workload (Cloudlet). This dataset is a set of resource utilization traces from PlanetLab VMs collected during 10 random days in March and April 2011. To find the experiment statistics, 500-3000 different workloads are executed. We selected the Poisson Distribution [36] for workload submission in this research work due to following reasons: 1) evaluating the performance of workload execution for specific interval of time and 2) every workload is independent of all other workloads (number of workloads are arriving in first hour is independent of the number of workloads arriving in any other hour).

Table 4: Simulation Parameters and their Values

| Parameter | Value |
|---|---|
| Number of VMs (nodes) | 360 |
| Number of Cloudlets (Workloads) | 3000 |
| Bandwidth | 1000 - 3000 B/S |
| CPU MIPS | 2000 |
| Size of Cloud Workload | 10000+ (10%–30%) MB |
| Number of PEs per machine | 1 |
| PE ratings | 100-4000 MIPS |
| Cost per Cloud Workload | 3 C\$–5 C\$ |
| Memory Size | 2048-12576 MB |
| File size | 300 + (15%–40%) MB |
| Cloud Workload output size | 300 + (15%–50%) MB |
| Initial Temperature | 12-22 °C |
| Inlet Temperature | 15-40 °C |
| Power Consumption by Processor | 130W – 240W |
| Power Consumption by Cooling Devices | 400 W – 900W |
| Power Consumption by RAM | 10W – 30W |
| Power Consumption by Storage | 35W – 110W |
| Power Consumption by Network | 70W-180W |
| Power Consumption by Extra Components | 2W-25W |

---

[2] PlanetLab - https://wikitech.wikimedia.org/wiki/Analytics/Archive/Data/Pagecounts-raw

### 4.3 Validation of ThermoSim

We have validated the proposed ThermoSim framework using an existing Thermal-aware Simulator, called **ThaS** [4], which implements the thermal-aware scheduling policy using CloudSim toolkit [7].

### 4.3.1 Baseline Simulator

ThaS [4] implements an interface between the hypervisor and the virtual machine in CDC. They have replaced the CloudSim's VMScheduler class which results in their scheduling behavior to be highly data-dependent of the existing class inputs. However, Thas lacks thermal characteristic studies by which they are not able to find the best host location at the time of VM migration. Moreover, their model is restricted in terms of analyzing the CPU and memory loads and the number of VM migrations which significantly impacts the performance of the system in terms of its latency and other QoS parameters. In ThermoSim, we have implemented the thermal-behaviour by developing three new classes (ThermalAware, UtilizationBased and TemperaturePredictor), which further extends the four classes (DataCenterCharacteristics, VMScheduler, VMAllocationPolicy and CloudletScheduler) of the CloudSim toolkit [7] through a coordinator class, *ThermoCloud*, as shown in Figure 6. The ThermalAware class takes temperature input from the TemperaturePredictor class when temperature information of the cloud hosts is not available. The TemperaturePredictor class uses utilization metrics from the UtilizationBased class and applies Fan-RPM models if such information is unavailable. We provided standard error bars for every graph to show the variation in the experimental results.

**Experimental Results:** Figure 8 (a) shows the variation of memory usage with varying numbers of workloads. ThermoSim saves 12.6% memory as compared to Thas, which increases further to 14.74% when using the prediction module. Figure 8 (b) shows the variation of time against the number of workloads. ThermoSim saves 17.7% of time as compared to Thas, which increases further to 23.64% when using the prediction module. Figure 8 (c) shows the variation of cost with number of workloads which is measured in Cloud Dollars (C$) and ThermoSim saves 15.5% cost as compared to Thas, which increases further to 20.45% when using the prediction module. Figure 8 (d) shows variation of energy with the number of workloads and ThermoSim saves 10.5% energy as compared to Thas, which increases further to 14.13% when using the prediction module.

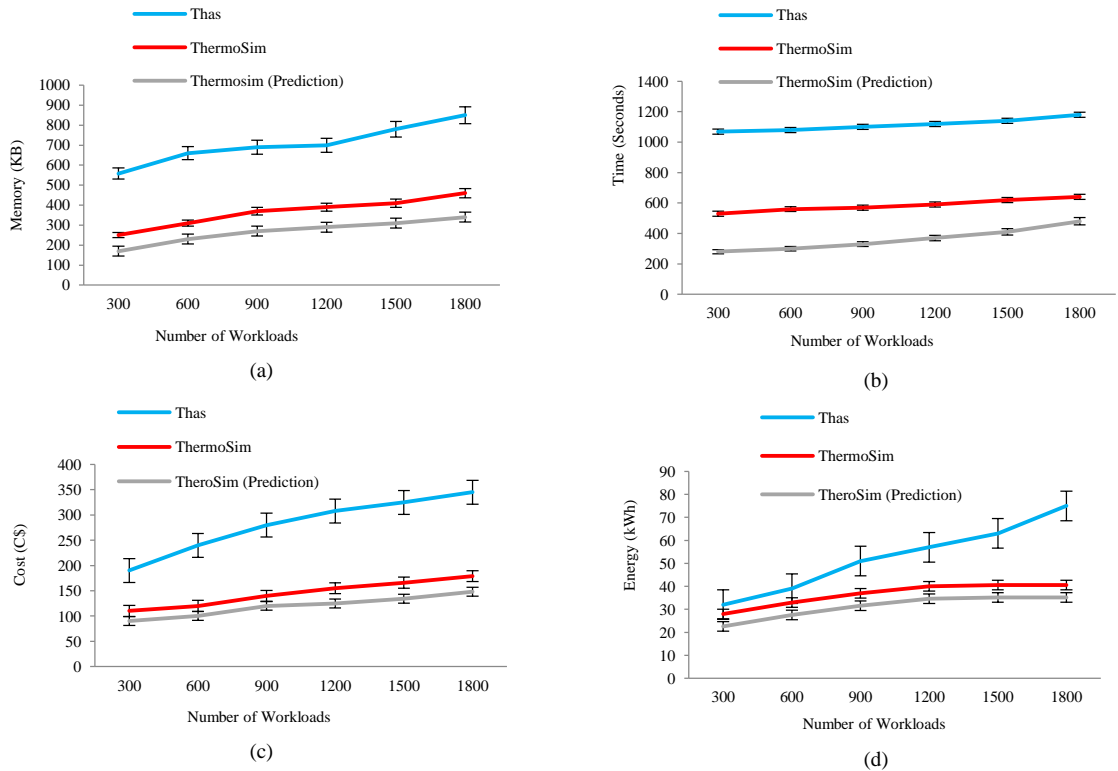

**Fig. 8.** Performance comparison of ThermoSim and Thas with different Number of Workloads: (a) Memory (b) Time, (c) Cost, (d) Energy

### 4.3.2 Prediction Accuracy

We have evaluated the prediction accuracy for both ThermoSim and Thas to prove the novelty of proposed framework with variation of workloads and nodes. *Prediction Accuracy is defined as the ratio of the number of correct predictions in the experiment to the total number of predictions in the experiment for thermal, energy and utilization.*

Figure 9 shows the comparison of ThermoSim and Thas based on prediction accuracy for resource requirement. Figure 9(a) shows the performance comparison of ThermoSim and Thas with different number of workloads and the value of prediction accuracy is decreasing with the increase in the number of workloads but ThermoSim performs better than Thas (both with and without the prediction module). Figure 9(b) shows the performance comparison of ThermoSim and Thas with different number of nodes and the value of prediction accuracy is increasing with the increase in the number of nodes but ThermoSim performs better than Thas (both with and without the prediction module).

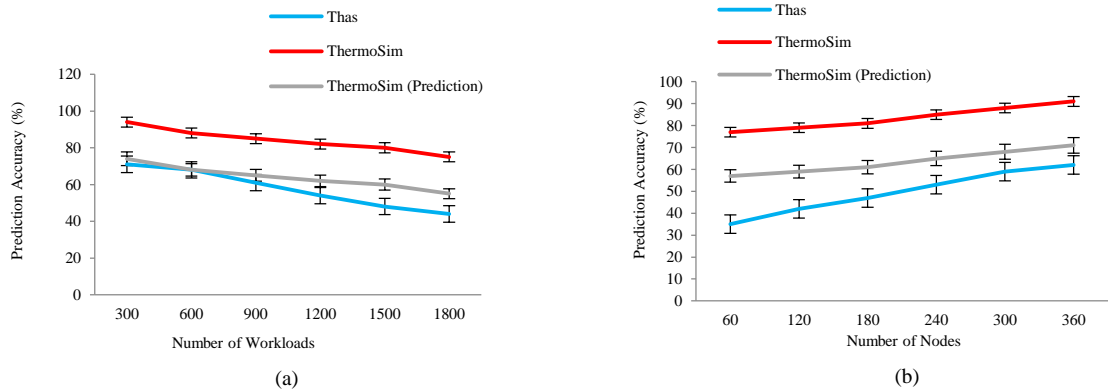

**Fig. 9.** Performance comparison of ThermoSim and Thas in terms of Prediction Accuracy for Resource Requirement: a) Different Number of Workloads and b) Different Number of Nodes

Figure 10 shows the comparison of ThermoSim and Thas based on prediction accuracy for temperature. Here too, the prediction accuracy decreases with increasing number of workloads and increases with increasing number of nodes. Figure 10 shows that the ThermoSim (both with and without prediction module) gives better performance as compared to Thas.

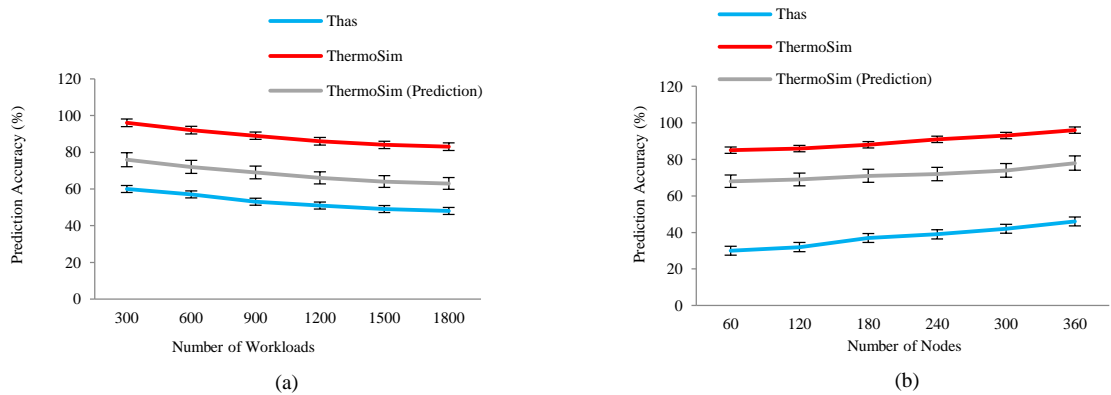

**Fig. 10.** Performance comparison of ThermoSim and Thas in terms of Prediction Accuracy for Temperature: a) Different Number of Workloads and b) Different Number of Nodes

Figure 11 shows the comparison of ThermoSim and Thas based on prediction accuracy for energy consumption. Figure 11(a) shows that the prediction accuracy falls drastically for Thas from 85% to 38% as number of workloads increases from 300 to 1800. Accuracy increases with number of nodes for Thas and ThermoSim (with and without the prediction module) as shown in Figure 11(b).

Figure 12 shows the comparison of ThermoSim and Thas based on prediction accuracy for resource utilization. Again, ThermoSim has higher accuracy compared to Thas. However, when the prediction module is used, the accuracy falls drastically with increasing number of workloads due to higher variance in memory usage characteristics at large number of workloads leading to thermal throttling in systems, hence reducing prediction performance [54].

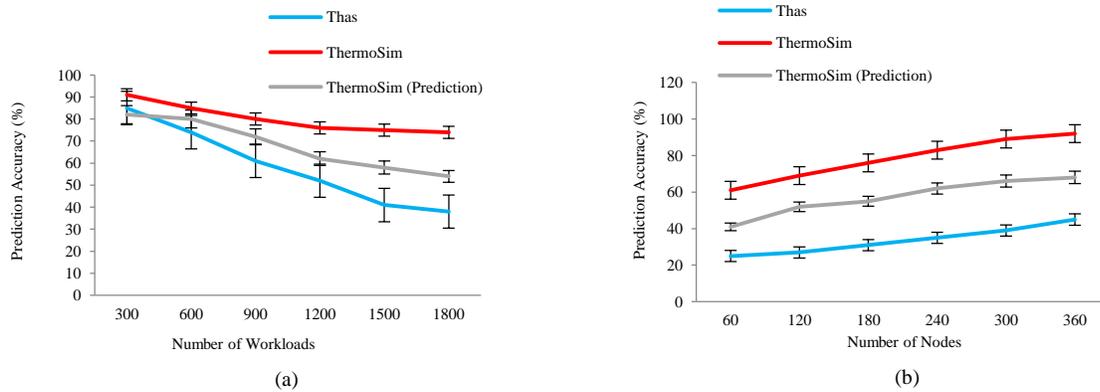

**Fig. 11.** Performance comparison of ThermoSim and Thas in terms of Prediction Accuracy for Energy Consumption: a) Different Number of Workloads and b) Different Number of Nodes

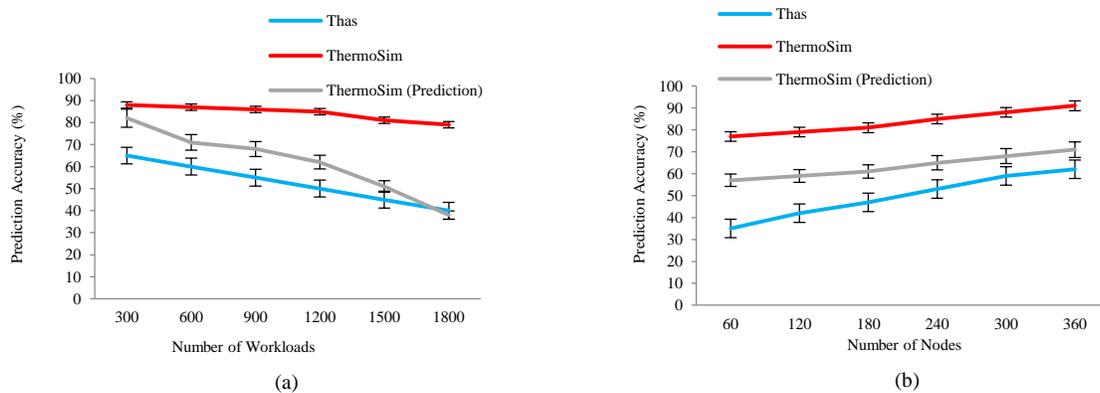

**Fig. 12.** Performance comparison of ThermoSim and Thas in terms of Prediction Accuracy for Resource Utilization: a) Different Number of Workloads and b) Different Number of Nodes

Figure 13 shows the comparison of ThermoSim and Thas based on prediction accuracy for memory utilization. Again, accuracy for ThermoSim (both with and without prediction module) is higher compared to Thas.

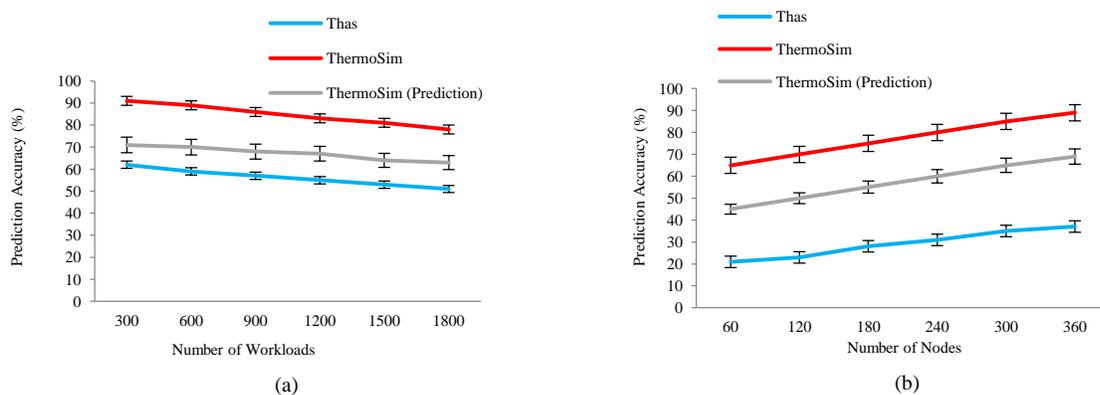

**Fig. 13.** Performance comparison of ThermoSim and Thas in terms of Prediction Accuracy for Memory Utilization: a) Different Number of Workloads and b) Different Number of Nodes

Figure 14 shows the comparison of ThermoSim and Thas based on prediction accuracy for disk utilization. Again, accuracy for ThermoSim (both with and without prediction module) is higher as compared to Thas. However, disk utilization prediction accuracy is low when using prediction modules due to high variance and task heterogeneity in the system [55].

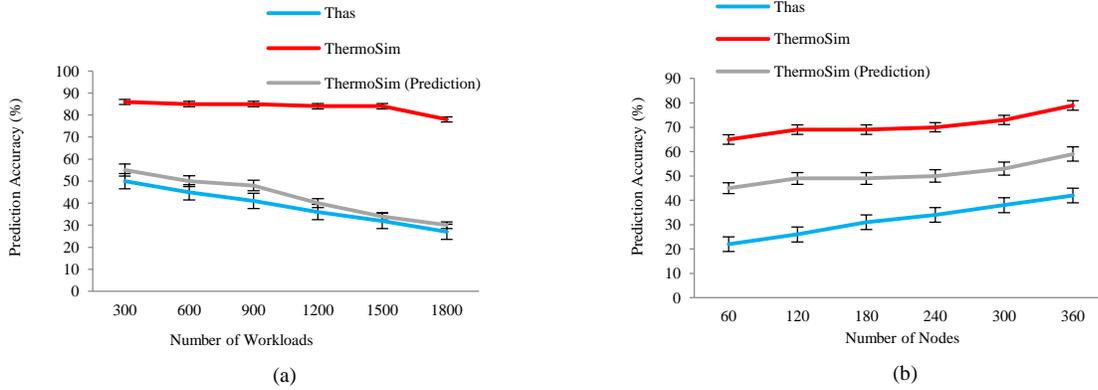

(a)                                                              (b)

**Fig. 14.** Performance comparison of ThermoSim and Thas in terms of Prediction Accuracy for Disk Utilization: a) Different Number of Workloads and b) Different Number of Nodes

Figure 15 shows the comparison of ThermoSim and Thas based on prediction accuracy for network utilization. As shown, the network utilization prediction accuracy decreases with increasing number of workloads and increases with increasing number of nodes.

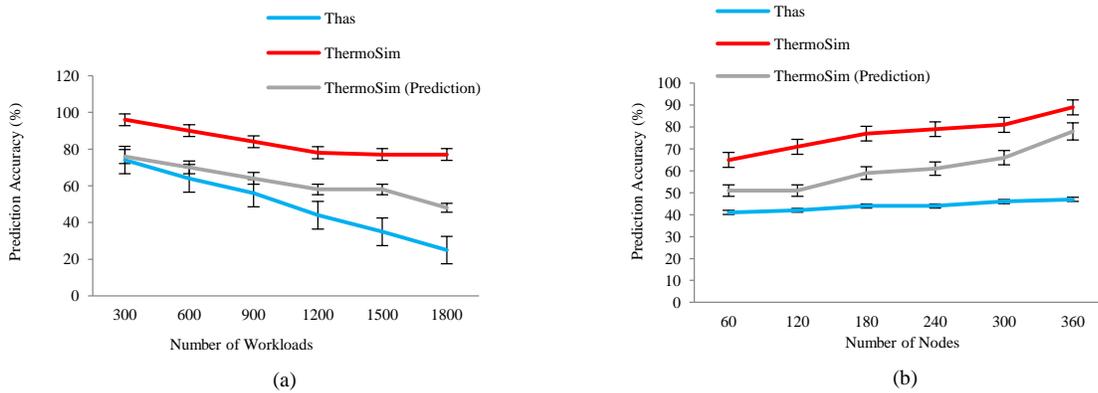

(a)                                                              (b)

**Fig. 15.** Performance comparison of ThermoSim and Thas in terms of Prediction Accuracy for Network Utilization: a) Different Number of Workloads and b) Different Number of Nodes

### 4.3.3 Analysis of Results

The ThermoSim framework is capable of scheduling the VMs independently. Experimental results show that ThermoSim has better performance compared to Thas as Thas implements thermal-aware scheduling policy inside the VMScheduler class of CloudSim toolkit [7], and as a result its scheduling behaviour is still dependent on the VMScheduler class - which increases data-dependency. There are two other important reasons for better performance of ThermoSim as compared to Thas: 1) ThermoSim is effective in locating and scheduling the energy efficient resources dynamically using CRUZE [18] and 2) ThermoSim performs scaling operations sharply. The prediction accuracy reduces when we use the prediction module as it does not consider various room cooling methods like CRAC, however due to much simpler implementation and lower computational requirements of the prediction module. ***Note:*** The detailed description all the metrics are given in our previous work [3] [18].

### 4.4 Case Studies using ThermoSim

We have presented two case studies to test the performance of three different energy-aware and thermal-aware resource management techniques using both the proposed ThermoSim framework and the existing framework Thas. The energy consumption shown here is based on average energies caluclated by the linear and non-linear models. For brevity and fair comparison, we only include results without using the predition module and using the Thermal model instead. The results when using the RNN are nearly between the ThermoSim and Thas results, as its prediction accuracies are lower than when Thermal model is used. This experiment used 1200 workloads. We have tested the performance of ThermoSim and Thas using QoS parameters such as energy consumption, SLA Violation Rate (SVR), number of VM migrations and temperature as shown in Figure 16 and Figure 17.

### 4.4.1 Case Study 1: Energy-aware Resource Management

In this section, first case study has been presented to test the performance of three different energy-aware resource management techniques (FCFS, DVFS and SOCCER) using both the proposed ThermoSim framework and the existing framework Thas [4]. First Come First Serve (FCFS) based energy-aware resource management technique schedules the resources for execution of homogeneous workloads using FCFS-based scheduling algorithm. Dynamic Voltage and Frequency Scaling (DVFS) [1] is an energy optimization approach, which adjusts the frequency settings of the computing devices to optimize scheduling of resources. SOCCER [3] is an energy-aware autonomic resource scheduling approach, which executes the heterogeneous cloud workloads using the IBM autonomic model. Figure 16 shows the variation of QoS parameters for different energy-aware resource management techniques.

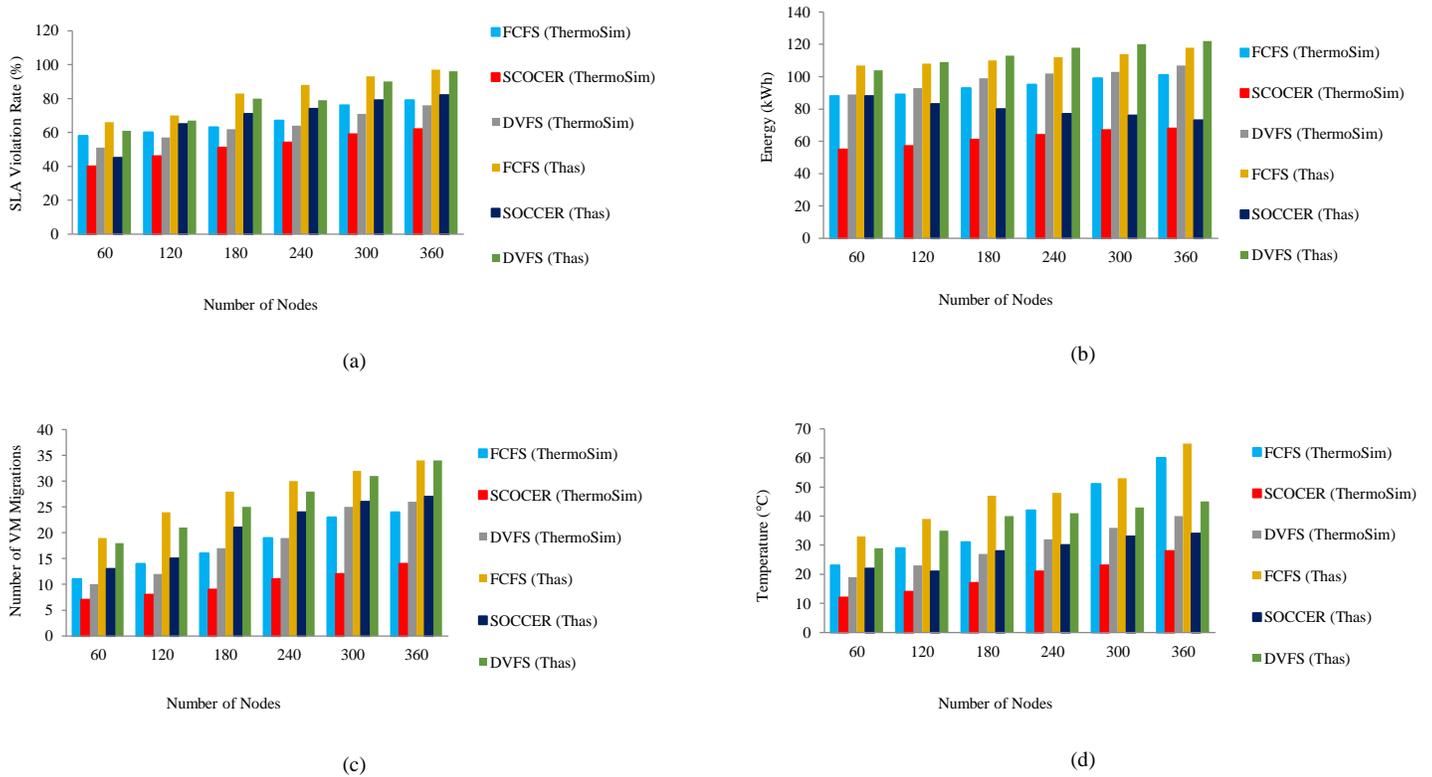

**Fig. 16.** Performance of different energy-aware scheduling algorithms: (a) SLA Violation Rate (b) Energy Consumption, (c) Number of VM Migrations, (d) Temperature

The value of QoS parameters increases as the number of nodes increases. Figure 16 (a) shows SOCCER has 13.12% less SLA violation rate in ThermoSim compared to Thas, similarly in FCFS and DVFS 12.45% and 14.98% improvement respectively. SOCCER performed better than DVFS and FCFS in both ThermoSim and Thas framework as it executes workloads based on signed SLA between user and provider. Figure 16 (b) shows SOCCER consumes 11.23% less energy in ThermoSim compared to Thas, similarly in FCFS and DVFS 8.35% and 9.15% improvement respectively. SOCCER performs better than DVFS and FCFS in both ThermoSim and Thas framework as it adjusts resource utilization at runtime.

Figure 16 (c) shows SOCCER has 10.45% fewer VM migrations in ThermoSim compared to Thas, similarly in FCFS and DVFS 5.15% and 5.91% improvement respectively. SOCCER perform better than DVFS and FCFS in both ThermoSim and Thas framework as it is capable of performing resource consolidation dynamically. Figure 16 (d) shows SOCCER offers 12.52% lower temperatures in ThermoSim compared to Thas, similarly in FCFS and DVFS 8.75% and 10.66% improvement respectively. SOCCER performs better than DVFS and FCFS in both ThermoSim and Thas framework as it shuts down the idle resources automatically during the execution of workloads.

### 4.4.2 Case Study 2: Thermal-aware Resource Management

In this section, second case study has been presented to test the performance of three different thermal-aware resource management techniques (DTM, ETAS and GTARA) using both the proposed ThermoSim framework and the existing framework Thas [4]. Energy and Thermal-Aware Scheduling (ETAS) algorithm [15] that dynamically consolidates VMs to minimize the overall energy consumption while proactively preventing hotspots. Dynamic Thermal Management (DTM) technique [51] exploits external computing resources (idle servers) adaptively as well as internal computing resources (free cores of CPU in the server) available in heterogeneous data centers. Game-based Thermal-Aware Resource Allocation (GTARA) strategy [13] decreases the imbalance within the CDC by using the concept of cooperative game theory with a Nash-bargaining to assign resources based on thermal profile. Figure 17 shows the variation of QoS parameters for different thermal-aware resource management techniques.

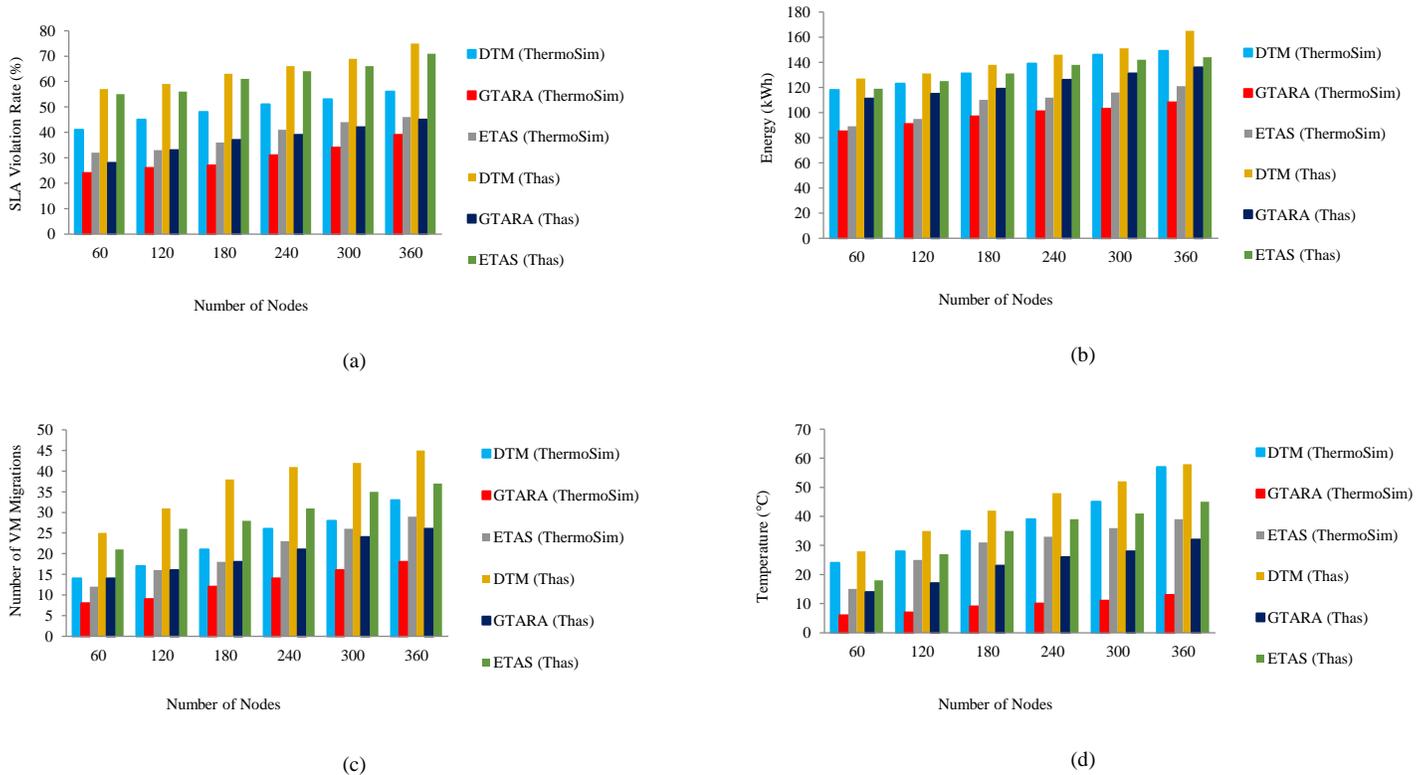

**Fig. 17.** Performance of different thermal-aware scheduling algorithms: (a) SLA Violation Rate (b) Energy Consumption, (c) Number of VM Migrations, (d) Temperature

The value of QoS parameters increases as the number of nodes increases. Figure 17 (a) shows GTARA has 15.32% less SLA violation rate in ThermoSim compared to Thas, similarly in DTM and ETAS 16.66% and 13.72% improvement respectively. GTARA performs better than ETAS and DTM in both ThermoSim and Thas framework as it reduces the energy consumption significantly with low VM migrations while not violating the SLAs. Figure 17 (b) shows GTARA consumes 9.73% less energy in ThermoSim compared to Thas, similarly in DTM and ETAS 11.93% and 12.65% improvement respectively. GTARA performs better than DTM and ETAS in both ThermoSim and Thas framework as it considers ambient effect of surrounding nodes is considered while assigning workloads to the computing nodes.

Figure 17 (c) shows GTARA has 10.87% fewer VM migrations in ThermoSim compared to Thas, similarly in DTM and ETAS 6.46% and 7.75% improvement respectively. GTARA perform better than ETAS and DTM in both ThermoSim and Thas framework as it can improve thermal balance and avoid hotspots dynamically by using cooperative game theory with a Nash-bargaining solution. Figure 17 (d) shows GTARA offers 18.65% lower temperatures in ThermoSim compared to Thas, similarly in DTM and ETAS 16.95% and 17.75% improvement respectively. GTARA perform better than DTM and ETAS in both ThermoSim and Thas framework because workloads are assigned to the nodes based on their thermal profiles.

**4.5 Discussions**

The experimental results demonstrate that ThermoSim is performing better than Thas in terms of different performance parameters such as memory, energy, temperature and cost. Further, it is clearly noted that the prediction accuracy of ThermoSim is better than Thas with the variation of number of workloads and nodes. The more holistic approach of ThermoSim consideration of diverse parameters like SLA violation rate, cost, number of VM migrations and energy independently for each VM allows it to surpass Thas in performance. As per the results of case study 1 using ThermoSim, SOCCER has better performance as compared to DVFS and FCFS using different QoS parameters for energy-aware techniques, while GTARA perform better than ETAS and DTM for thermal-aware techniques as shown in case study 2.

# 5 Summary and Conclusions

In this paper, we proposed a framework called ThermoSim for the simulation and modelling of thermal-aware resource management for cloud computing environments. ThermoSim uses a lightweight RNN-based deep leaning model for thermal-predictor for efficient and low overhead resource management in resource constrained cloud environments. We have validated the proposed ThermoSim framework using three well-known energy-aware and thermal-aware resource scheduling techniques by testing the performance of four QoS parameters: energy consumption, SLA violation rate, number of VM migrations and temperature, with different number of resources. We have validated the proposed ThermoSim framework against existing thermal-aware simulator (Thas). Finally, relationship between theory and practice is very important. Benchmarking is an important starting point, which may try to relate the holistic aspects studied in our simulation in real-world practice. This may lead to additional improvements of the theoretical basis.

**5.1 Future Research Directions and Open Challenges**

Although the ThermoSim framework is capable of simulating and modeling the thermal-aware characteristics of cloud data centers, it can be further enhanced in a larger scope under the following aspects.

1. *High Energy Demand in Cooling Servers*: High power consumption leads to the creation of hot spots and increases in server temperature. Thus, there is a need to study the trade-off between cooling energy and computing energy and its effect on temperature.

2. *Peak Temperature among Servers*: Temperature is another important parameter for both physical servers and virtualization solutions. Variance in the on-chip temperature and the resultant occurrence of hot spots degrades the performance of processors, increases the energy consumption. Thermal management strategies are required to uniformly distribute the temperature.

3. *High Level of Power Consumption by the Servers*: The non-energy aware scheduling techniques lead to increase in power consumption among the servers which degrades the server's reliability and performance in terms of availability and scalability.

4. *Validation of ThermoSim in the Large-scale Cloud Data Center*: The proposed ThermoSim framework will be implemented in a real cloud environment to validate the availability of the proposed models using Hadoop based Cloud Cluster and thermal sensors.

5. *Automation of ThermoSim using Artificial Intelligence (AI):* To build a framework/benchmark that automatically classifies applications/workloads according to their temperature profile using AI [46].

6. *Fog and Edge Computing Environments:* Further, ThermoSim can be extended towards fog/edge computing scenarios with heterogeneous hardware, and stronger energy constraints [46].

7. *Container-based Deployment:* In future, virtualization technology (e.g., VMs) used in ThermoSim can be replaced with Docker-based containers to improve the performance for CPU-intensive applications. Ease of use of containers (especially quick restarts) can reduce the execution time of workloads and improves energy efficiency [45]. Containers also provide a lightweight environment for the deployment applications because they are stand-alone, self-contained units that package software and its dependencies together. Further, container-based deployment is more effective than VMs because containers have small memory footprint and consume a very small amount of resources [46].

8. *Reliability and Security:* There is a need to investigate the interlink between cloud thermo properties and dependable provision of computation, covering scheduling, resource allocation, consolidation, execution, cloudlet among the others etc. By dependability, we refer to the extent to which the cloud environment can continue to maintain its reliability, security and performance while dynamically optimizing for energy and honouring thermo-properties [46].

9. *Dynamic Monitoring:* There is a need to develop priority tools and repositories for monitoring and collecting behavioural information about the dynamic management of thermo proprieties and dependability. These repositories can be mined to inform the design of more dependable thermo-aware policies.

**Declaration of Interests**

The authors declare that they have no known competing financial interests or personal relationships that could have appeared to influence the work reported in this paper.

**CRediT authorship contribution statement**

**Sukhpal Singh Gill:** Conceptualization, Data curation, Formal analysis, Investigation, Methodology, Software, Visualization, Validation, Writing - original draft and Writing - review & editing. **Shreshth Tuli:** Conceptualization, Data curation, Methodology, Software, Validation and Writing - review & editing. **Adel Nadjaran Toosi:** Writing - review & editing, Data curation and Supervision. **Felix Cuadrado:** Formal analysis, Software and Writing - review & editing. **Peter Garraghan:** Writing -original draft, Formal analysis, Supervision, Methodology and Writing - review & editing. **Rami Bahsoon:** Data curation and Writing - review & editing. **Hanan Lutfiyya:** Visualization and Writing - review & editing. **Rizos Sakellariou:** Formal analysis and Writing - review & editing. **Omer Rana:** Conceptualization, Supervision and Writing - review & editing. **Schahram Dustdar:** Visualization and Writing - review & editing. **Rajkumar Buyya:** Conceptualization, Formal analysis, Writing -original draft, Supervision, Visualization, Methodology and Writing - review & editing.

**Datasets**

The datasets used for this research work are available at:

Alibaba Cluster - https://github.com/alibaba/clusterdata

PlanetLab - https://wikitech.wikimedia.org/wiki/Analytics/Archive/Data/Pagecounts-raw

**Acknowledgements**

We would like to thank Joseph Richardson (Lancaster University, UK), Shashikant Ilager (University of Melbourne, Australia), Shikhar Tuli (IIT Delhi, India) and Dominic Lindsay (Lancaster University, UK) for their feedback to improve the quality of the paper. We would like to thank the editors, area editor and anonymous reviewers for their valuable comments and suggestions to help and improve our research paper. An initial investigation on this work was carried out at Melbourne CLOUDS Lab, which was supported by Melbourne-Chindia Cloud Computing (MC3) Research Network.

## References

1. Chia-Ming W, Chang R., and Chan H.: A green energy-efficient scheduling algorithm using the DVFS technique for cloud datacenters. Future Generation Computer Systems 37, 141-147 (2014).
2. Tsai, Ting-Hao, and Ya-Shu Chen. "Thermal-throttling server: A thermal-aware real-time task scheduling framework for three-dimensional multicore chips." Journal of Systems and Software 112 (2016): 11-25.
3. Singh S., Chana I, Singh M., and Buyya R.: SOCCER: self-optimization of energy-efficient cloud resources. Cluster Computing, 19, no. 4, 1787-1800 (2016).
4. Mhedheb Y., Jrad F., Tao J., Zhao J., Kołodziej J., and Streit A.: Load and thermal-aware VM scheduling on the cloud. In the Proceeding of International Conference on Algorithms and Architectures for Parallel Processing, 101-114. Springer, Cham (2013)
5. Chu, Hsin-Hao, Yu-Chon Kao, and Ya-Shu Chen. "Adaptive thermal-aware task scheduling for multi-core systems." Journal of Systems and Software 99 (2015): 155-174
6. Zhou, Junlong, and Tongquan Wei. "Stochastic thermal-aware real-time task scheduling with considerations of soft errors." Journal of Systems and Software 102 (2015): 123-133
7. Calheiros R.N., Ranjan R., Beloglazov A., Rose C.A.F.D., and Buyya R.: CloudSim: a toolkit for modeling and simulation of cloud computing environments and evaluation of resource provisioning algorithms. Software: Practice and Experience, 41, no. 1, 23-50 (2011).
8. Xiang L., Garraghan P., Jiang X., Wu Z., and Xu J.: Holistic virtual machine scheduling in cloud datacenters towards minimizing total energy., IEEE Transactions on Parallel and Distributed Systems 29, no. 6, 1317-1331 (2018).
9. Wu, Guowei, and Zichuan Xu. "Temperature-aware task scheduling algorithm for soft real-time multi-core systems." Journal of Systems and Software 83, no. 12 (2010): 2579-2590
10. Rodero, Ivan, Hariharasudhan Viswanathan, Eun Kyung Lee, Marc Gamell, Dario Pompili, and Manish Parashar. "Energy-efficient thermal-aware autonomic management of virtualized HPC cloud infrastructure." Journal of Grid Computing 10, no. 3 (2012): 447-473.
11. Kumar, Mohan Raj Velayudhan, and Shriram Raghunathan. "Heterogeneity and thermal aware adaptive heuristics for energy efficient consolidation of virtual machines in infrastructure clouds." Journal of Computer and System Sciences 82, no. 2 (2016): 191-212.
12. Liu, Huazhong, Baoshun Liu, Laurence T. Yang, Man Lin, Yuhui Deng, Kashif Bilal, and Samee U. Khan. "Thermal-aware and DVFS-enabled big data task scheduling for data centers." IEEE Transactions on Big Data 4, no. 2 (2018): 177-190.
13. Akbar, Saeed, Saif Ur Rehman Malik, Samee U. Khan, Raymond Choo, Adeel Anjum, and Naveed Ahmad. "A Game-based Thermal-aware Resource Allocation Strategy for Data Centers." IEEE Transactions on Cloud Computing (2019).
14. Khaleel, Mustafa I. "Load Balancing and Thermal-Aware in Geo-Distributed Cloud Data Centers Based on Vlans." Science Journal of University of Zakho 6, no. 3 (2018): 112-117.
15. Ilager, Shashikant, Kotagiri Ramamohanarao, and Rajkumar Buyya. "ETAS: Energy and thermal-aware dynamic virtual machine consolidation in cloud data center with proactive hotspot mitigation." Concurrency and Computation: Practice and Experience 31, no. 17 (2019): e5221.
16. Capozzoli A., and Primiceri G.: Cooling systems in data centers: state of art and emerging technologies. Energy Procedia, 83, 484-493 (2015).
17. Lee E. K., Viswanathan H., and Pompili D.: Proactive thermal-aware resource management in virtualized HPC cloud datacenters. IEEE Transactions on Cloud Computing, 5, no. 2, 234-248 (2017).
18. Sukhpal Singh Gill, Peter Garraghan, Vlado Stankovski, Giuliano Casale, Soumya K. Ghosh, Ruppa K. Thulasiram, Kotagiri Ramamohanarao and Rajkumar Buyya, "Holistic Resource Management for Sustainable and Reliable Cloud Computing: An Innovative Solution to Global Challenge", Journal of Systems and Software, Elsevier, Volume 155, Pages: 104-129, 2019.
19. Moore, Justin D., Jeffrey S. Chase, Parthasarathy Ranganathan, and Ratnesh K. Sharma. "Making Scheduling "Cool": Temperature-Aware Workload Placement in Data C Zhang S, Chatha K S. Approximation Algorithm for the Temperature Aware Scheduling Problem. Proceedings of International Conference on Computer-Aided Design. 2007:281–288.
20. Tang Qinghui, Gupta Sandeep Kumar S, Varsamopoulos Georgios. Energy-efficient thermal-aware task scheduling for homogeneous highperformance computing data centers: A cyber-physical approach. IEEE Transactions on Parallel and Distributed Systems. 2008;19(11):1458–1472
21. Lazic, Nevena, Craig Boutilier, Tyler Lu, Eehern Wong, Binz Roy, M. K. Ryu, and Greg Imwalle. "Data center cooling using model-predictive control." In Advances in Neural Information Processing Systems, pp. 3814-3823. 2018.
22. Parthasarathy Ranganathan, and Ratnesh K. Sharma. "Making Scheduling" Cool": Temperature-Aware Workload Placement in Data Centers." In USENIX annual technical conference, General Track, pp. 61-75. 2005.
23. Chaudhry M. T., Ling T. C., Manzoor A., Hussain S. A., and Kim J.: Thermal-aware scheduling in green data centers. ACM Computing Surveys 47, no. 3, 1-39 (2015)
24. Lin, Weiwei, Guangxin Wu, Xinyang Wang, and Keqin Li. "An artificial neural network approach to power consumption model construction for servers in cloud data centers." IEEE Transactions on Sustainable Computing (2019).
25. Xu, Minxian, and Rajkumar Buyya. "BrownoutCon: A Software System based on Brownout and Containers for Energy-Efficient Cloud Computing." Journal of Systems and Software, Volume 155, September 2019, Pages 91-103.
26. Singh, Sukhpal, and Inderveer Chana. "A survey on resource scheduling in cloud computing: Issues and challenges." Journal of grid computing 14, no. 2 (2016): 217-264.
27. Zhou, Junlong, Kun Cao, Peijin Cong, Tongquan Wei, Mingsong Chen, Gongxuan Zhang, Jianming Yan, and Yue Ma. "Reliability and temperature constrained task scheduling for makespan minimization on heterogeneous multi-core platforms." Journal of Systems and Software 133 (2017): 1-16.
28. Grozev, Nikolay, and Rajkumar Buyya. "Performance modelling and simulation of three-tier applications in cloud and multi-cloud environments." The Computer Journal 58, no. 1 (2013): 1-22.
29. Lebre, Adrien, Arnaud Legrand, Frédéric Suter, and Pierre Veyre. "Adding storage simulation capacities to the simgrid toolkit: Concepts, models, and api." In 2015 15th IEEE/ACM International Symposium on Cluster, Cloud and Grid Computing, pp. 251-260. IEEE, 2015.
30. Kouki, Yousri, and Thomas Ledoux. "Sla-driven capacity planning for cloud applications." In Cloud Computing Technology and Science (CloudCom), 2012 IEEE 4th International Conference on, pp. 135-140. IEEE, 2012.
31. Möbius, Christoph, Waltenegus Dargie, and Alexander Schill. "Power consumption estimation models for processors, virtual machines, and servers." IEEE Transactions on Parallel and Distributed Systems 25, no. 6 (2014): 1600-1614

32. Li, Xiang, Xiaohong Jiang, Peter Garraghan, and Zhaohui Wu. "Holistic energy and failure aware workload scheduling in Cloud datacenters." Future Generation Computer Systems 78 (2018): 887-900.

33. Balis, Bartosz, Robert Brzoza-Woch, Marian Bubak, Marek Kasztelnik, Bartosz Kwolek, Piotr Nawrocki, Piotr Nowakowski, Tomasz Szydlo, and Krzysztof Zielinski. "Holistic approach to management of IT infrastructure for environmental monitoring and decision support systems with urgent computing capabilities." Future Generation Computer Systems 79 (2018): 128-143.

34. Liu, Zhenhua, Yuan Chen, Cullen Bash, Adam Wierman, Daniel Gmach, Zhikui Wang, Manish Marwah, and Chris Hyser. "Renewable and cooling aware workload management for sustainable data centers." In ACM SIGMETRICS Performance Evaluation Review, vol. 40, no. 1, pp. 175-186. ACM, 2012.

35. Wu, Wentai, Weiwei Lin, Ligang He, Guangxin Wu, and Ching-Hsien Hsu. "A Power Consumption Model for Cloud Servers Based on Elman Neural Network." IEEE Transactions on Cloud Computing (2019).

36. Amvrosiadis, George, Jun Woo Park, Gregory R. Ganger, Garth A. Gibson, Elisabeth Baseman, and Nathan DeBardeleben. "On the diversity of cluster workloads and its impact on research results." In 2018 {USENIX} Annual Technical Conference ({USENIX} {ATC} 18), pp. 533-546. 2018.

37. Lin, Weiwei, Gaofeng Peng, Xinran Bian, Siyao Xu, Victor Chang, and Yin Li. "Scheduling Algorithms for Heterogeneous Cloud Environment: Main Resource Load Balancing Algorithm and Time Balancing Algorithm." Journal of Grid Computing 17, no. 4 (2019): 699-726.

38. Lin, Weiwei, Wentai Wu, and Ligang He. "An On-line Virtual Machine Consolidation Strategy for Dual Improvement in Performance and Energy Conservation of Server Clusters in Cloud Data Centers." IEEE Transactions on Services Computing (2019).

39. Cho, Kyunghyun; van Merrienboer, Bart; Gulcehre, Caglar; Bahdanau, Dzmitry; Bougares, Fethi; Schwenk, Holger; Bengio, Yoshua (2014). "Learning Phrase Representations using RNN Encoder-Decoder for Statistical Machine Translation". arXiv:1406.1078

40. Saha, Saraswati, and Anupam Majumdar. "Data centre temperature monitoring with ESP8266 based Wireless Sensor Network and cloud based dashboard with real time alert system." In 2017 Devices for Integrated Circuit (DevIC), pp. 307-310. IEEE, 2017.

41. Bakker, Anton, and Johan Huijsing. High-accuracy CMOS smart temperature sensors. Vol. 595. Springer Science & Business Media, 2013.

42. Dey, Rahul, and Fathi M. Salemt. "Gate-variants of gated recurrent unit (GRU) neural networks." In 2017 IEEE 60th international midwest symposium on circuits and systems (MWSCAS), pp. 1597-1600. IEEE, 2017.

43. Zhao, Hui, Jing Wang, Feng Liu, Quan Wang, Weizhan Zhang, and Qinghua Zheng. "Power-aware and performance-guaranteed virtual machine placement in the cloud." IEEE Transactions on Parallel and Distributed Systems 29, no. 6 (2018): 1385-1400.

44. K. Park and V. S. Pai, "Comon: a mostly-scalable monitoring system for planetlab," ACM SIGOPS Operating Systems Review, vol. 40, no. 1, pp. 65–74, 2006.

45. Zhou, Qiheng, Minxian Xu, Sukhpal Singh Gill, Chengxi Gao, Wenhong Tian, Chengzhong Xu, and Rajkumar Buyya. "Energy Efficient Algorithms based on VM Consolidation for Cloud Computing: Comparisons and Evaluations." In the Proceedings of the 20th IEEE/ACM International Symposium on Cluster, Cloud and Grid Computing (CCGRID 2020), 1-10, Online Available at: arXiv:2002.04860 (2020).

46. Gill, Sukhpal Singh, Shreshth Tuli, Minxian Xu, Inderpreet Singh, Karan Vijay Singh, Dominic Lindsay, Shikhar Tuli et al. Transformative effects of IoT, Blockchain and artificial intelligence on cloud computing: evolution, vision, trends and open challenges. Internet of Things. 2019; 8:100118.

47. Li, Fengcun, and Bo Hu. "DeepJS: Job Scheduling Based on Deep Reinforcement Learning in Cloud Data Center." In Proceedings of the 2019 4th International Conference on Big Data and Computing, pp. 48-53. 2019.

48. Guo, Jing, Zihao Chang, Sa Wang, Haiyang Ding, Yihui Feng, Liang Mao, and Yungang Bao. "Who limits the resource efficiency of my datacenter: an analysis of Alibaba datacenter traces." In Proceedings of the International Symposium on Quality of Service, pp. 1-10. 2019.

49. Tian, Huangshi, Yunchuan Zheng, and Wei Wang. "Characterizing and Synthesizing Task Dependencies of Data-Parallel Jobs in Alibaba Cloud." In Proceedings of the ACM Symposium on Cloud Computing, pp. 139-151. 2019.

50. Wu, Heng, Wenbo Zhang, Yuanjia Xu, Hao Xiang, Tao Huang, Haiyang Ding, and Zheng Zhang. "Aladdin: Optimized Maximum Flow Management for Shared Production Clusters." In 2019 IEEE International Parallel and Distributed Processing Symposium (IPDPS), pp. 696-707. IEEE, 2019.

51. Kim, Young Geun, Jeong In Kim, Seung Hun Choi, Seon Young Kim, and Sung Woo Chung. "Temperature-aware Adaptive VM Allocation in Heterogeneous Data Centers." In 2019 IEEE/ACM International Symposium on Low Power Electronics and Design (ISLPED), pp. 1-6. IEEE, 2019.

52. Fu, Lijun, Jianxiong Wan, Ting Liu, Xiang Gui, and Ran Zhang. "A temperature-aware resource management algorithm for holistic energy minimization in data centers." In 2017 2nd Workshop on Recent Trends in Telecommunications Research (RTTR), pp. 1-5. IEEE, 2017.

53. Dundu, Morgan, and Xiaojin Gao, eds. Construction Materials and Structures: Proceedings of the First International Conference on Construction Materials and Structures. IOS Press, 2014.

54. Lin, Jiang, Hongzhong Zheng, Zhichun Zhu, Howard David, and Zhao Zhang. "Thermal modeling and management of DRAM memory systems." In Proceedings of the 34th annual international symposium on Computer architecture, pp. 312-322. 2007.

55. Longford, Graham. Network neutrality'vs.'network diversity': A survey of the debate, policy landscape and implications for broadband as an essential service for Ontarians. Working paper for the Ministry of Government Services (Ontario). CRACIN, nd Web. 27 Jan, 2010.

56. Fu, Lijun, Jianxiong Wan, Jie Yang, Dongdong Cao, and Gefei Zhang. "Dynamic thermal and IT resource management strategies for data center energy minimization." Journal of Cloud Computing 6, no. 1 (2017): 1-16.

57. Chen, Ying-Jun, Gwo-Jiun Horng, Jian-Hua Li, and Sheng-Tzong Cheng. "Using Thermal-Aware VM Migration Mechanism for High-Availability Cloud Computing." Wireless Personal Communications 97, no. 1 (2017): 1475-1502.

58. Wu, Zhaohui, Xiang Li, Peter Garraghan, Xiaohong Jiang, Kejiang Ye, and Albert Y. Zomaya. "Virtual machine level temperature profiling and prediction in cloud datacenters." In 2016 IEEE 36th International Conference on Distributed Computing Systems (ICDCS), pp. 735-736. IEEE, 2016.

59. Oxley, Mark A., Sudeep Pasricha, Anthony A. Maciejewski, Howard Jay Siegel, and Patrick J. Burns. "Online resource management in thermal and energy constrained heterogeneous high performance computing." In 2016 IEEE 14th Intl Conf on Dependable, Autonomic and Secure Computing, 14th Intl Conf on Pervasive Intelligence and Computing, 2nd Intl Conf on Big Data Intelligence and Computing and Cyber Science and Technology Congress (DASC/PiCom/DataCom/CyberSciTech), pp. 604-611. IEEE, 2016.

60. Pierson, Jean-Marc, Patricia Stolf, Hongyang Sun, and Henri Casanova. "MILP formulations for spatio-temporal thermal-aware scheduling in Cloud and HPC datacenters." Cluster Computing (2019): 1-19.

**Author's Biography**

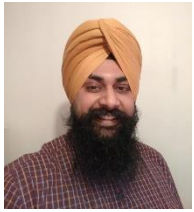

**Sukhpal Singh Gill** is a Lecturer (Assistant Professor) in Cloud Computing at School of Electronic Engineering and Computer Science (EECS), Queen Mary University of London, UK. Prior to this, Dr. Gill has held positions as a Research Associate at the School of Computing and Communications, Lancaster University, UK and also as a Postdoctoral Research Fellow at the Cloud Computing and Distributed Systems (CLOUDS) Laboratory, School of Computing and Information Systems, The University of Melbourne, Australia. Dr. Gill was a research visitor at Monash University, University of Manitoba and Imperial College London. He was a recipient of several awards, including the Distinguished Reviewer Award from Software: Practice and Experience (Wiley), 2018, and served as the PC member for venues such as UCC, SE-CLOUD, ICCCN, ICDICT and SCES. His one review paper has been nominated and selected for the ACM 21st annual Best of Computing Notable Books and Articles as one of the notable items published in computing - 2016. He has published 50+ papers as a leading author in highly ranked journals and conferences with H-index 20. Dr. Gill has reviewed 160+ research articles of high ranked journals and prestigious conferences. His research interests include Cloud Computing, Internet of Things, Energy Efficiency and Machine Learning. For further information on Dr. Gill, please visit: www.ssgill.me.

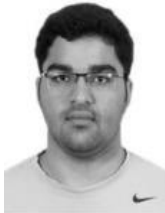

**Shreshth Tuli** is an undergraduate student at the Department of Computer Science and Engineering at Indian Institute of Technology - Delhi, India. He is a national level Kishore Vaigyanic Protsahan Yojana (KVPY) scholarship holder for excellence in science and innovation. He has worked as a visiting researcher at the Cloud Computing and Distributed Systems (CLOUDS) Laboratory, Department of Computing and Information Systems, the University of Melbourne, Australia. Most of his projects are focused on developing technologies for future requiring sophisticated hardware-software integration. He has published his research work in top tier journals and conferences, including ACM Transactions on Embedded Computing Systems, Elsevier JSS and FCGS, IEEE CloudCom and Wiley ITL. He is an active reviewer of SPE Wiley, Elsevier JSS and FCGS. His research interests include Internet of Things (IoT), Fog Computing, Network Design, and Artificial Intelligence.

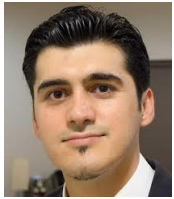

**Adel Nadjaran Toosi** is a Lecturer in computer systems at Faculty of Information Technology, Monash University, Australia. He received his B.sc. degree in 2003 and his M.Sc. degree in 2006 both in Computer Science and Software Engineering from Ferdowsi University of Mashhad, Iran and his Ph.D. degree in 2015 from the University of Melbourne. Adel's Ph.D. studies were supported by International Research Scholarship (MIRS) and Melbourne International Fee Remission Scholarship (MIFRS). His Ph.D. thesis was nominated for CORE John Makepeace Bennett Award for the Australasian Distinguished Doctoral Dissertation and John Melvin Memorial Scholarship for the Best Ph.D. thesis in Engineering. His research interests include scheduling and resource provisioning mechanisms for distributed systems. Currently, he is working on resource management in Software-Defined Networks (SDN)-enabled Cloud Computing.

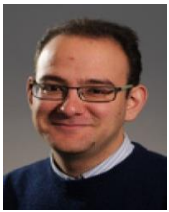

**Felix Cuadrado** is a Distinguished Academic (Beatriz Galindo Senior Fellowship) in the Technical University of Madrid (UPM) and a Fellow of the Alan Turing Institute. Prior to joining UPM, he was Senior Lecturer (Associate Professor) in the School of Electronic Engineering and Computer Science at Queen Mary University of London. His research explores the challenges arising from large-scale applications through a combination of software engineering, distributed systems, measurement, and mathematical approaches. He has numerous publications in top tier journals and conferences, including IEEE Transactions on Software Engineering, IEEE Transactions on Cloud Computing, Elsevier JSS and FCGS, IEEE ICDCS, and WWW. He also has substantial experience in funded research, being Co-PI for European projects (FP7 TRIDEC and H2020 ENDEAVOUR — Internet-scale network management), as well as the EPSRC project EARL (measurements at Internet scale) and Turing Project Raphtory.

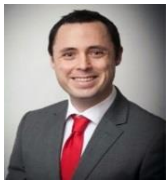

**Peter Garraghan** is a Lecturer in the School of Computing & Communications, Lancaster University. His primary research expertise is studying the complexity and emergent behaviour of massive-scale distributed systems (Cloud computing, Datacenters, Internet of Things) to propose design new techniques for enhancing system dependability, resource management, and energy-efficiency. Peter has industrial experience building large-scale production distributed systems and has worked and collaborated internationally with the likes of Alibaba Group, Microsoft, STFC, CONACYT, and the UK Datacenter and IoT industry.

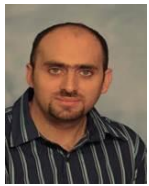

**Rami Bahsoon** is a Senior Lecturer (Associate Professor) at the School of Computer Science, University of Birmingham, UK. Bahsoon's research is in the area of software architecture, cloud and services software engineering, self-aware software architectures, self-adaptive and managed software engineering, economics-driven software engineering and technical debt management in software. He co-edited four books on Software Architecture, including Economics-Driven Software Architecture; Software Architecture for Big Data and the Cloud; Aligning Enterprise, System, and Software Architecture. He was a Visiting Scientist at the Software Engineering Institute (SEI), Carnegie Mellon University, USA (June-August 2018) and was the 2018 Melbourne School of Engineering (MSE) Visiting Fellow of The School of Computing and Information Systems, the University of Melbourne (August to Nov 2018). He holds a PhD in Software Engineering from University College London (2006) and was MBA Fellow in Technology at London Business School (2003–2005). He is a fellow of the Royal Society of Arts and Associate Editor of IEEE Software - Software Economies.

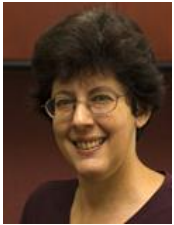

**Hanan Lutfiyya** is a Professor and the Chair of the Department of Computer Science, University of Western Ontario (UWO), Canada. Her research interests include Internet of Things, software engineering, self-adaptive and self-managing systems, autonomic computing, monitoring and diagnostics, mobile systems, policies, and clouds. She was a recipient of the UWO Faculty Scholar Award in 2006. She is a Past Member of the Natural Science and Engineering Research Council of Canada (NSERC) Discovery Grant Committee, and a Past Member and the Chair of an NSERC Strategic Grants Committee. She was a member of the Computer Science Accreditation Council. She is currently an Associate Editor of the IEEE Transactions on Network and Service Management and has recently served as the Program Co-Chair for the IEEE/IFIP Network Operations and Management Symposium and the IEEE International Conference on Network and Service Management. She is currently on the steering committee for the Ontario Celebration of Women in Computing Conference.

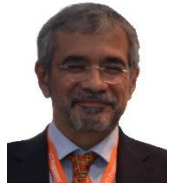

**Rizos Sakellariou** obtained his PhD from the University of Manchester in 1997. Since then he held positions with Rice University and the University of Cyprus, while currently he is with the University of Manchester leading a laboratory that carries out research in High-Performance, Parallel and Distributed systems, which over the last 10 years has hosted more than 30 doctoral students, researchers and long-term visitors. Rizos has carried out research on a number of topics related to parallel and distributed computing, with an emphasis on problems stemming from efficient resource utilization and workload allocation and scheduling issues. He has published over 140 research papers, His research has been supported by several UK and EU projects and has been on the Program Committee of over 150 conferences and workshops. He values collaboration and a strong work ethic.

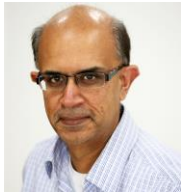

**Omer Rana** is a Professor of Performance Engineering in School of Computer Science & Informatics at Cardiff University and Deputy Director of the Welsh e-Science Centre. He holds a PhD from Imperial College. His research interests extend to three main areas within computer science: problem solving environments, high performance agent systems and novel algorithms for data analysis and management. Moreover, he leads the Complex Systems research group in the School of Computer Science & Informatics and is director of the "Internet of Things" Lab, at Cardiff University.

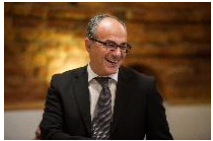

**Schahram Dustdar** is Full Professor of Computer Science heading the Research Division of Distributed Systems at the TU Wien, Austria. He holds several honorary positions: University of California (USC) Los Angeles; Monash University in Melbourne, Shanghai University, Macquarie University in Sydney, and University of Groningen (RuG), The Netherlands (2004-2010). From Dec 2016 until Jan 2017 he was a Visiting Professor at the University of Sevilla, Spain and from January until June 2017 he was a Visiting Professor at UC Berkeley, USA. From 1999 - 2007 he worked as the co-founder and chief scientist of Caramba Labs Software AG in Vienna (acquired by Engineering NetWorld AG), a venture capital co-funded software company focused on software for collaborative processes in teams. Caramba Labs was nominated for several (international and national) awards: World Technology Award in the category of Software (2001); Top-Startup companies in Austria (Cap Gemini Ernst & Young) (2002); MERCUR Innovation award of the Austrian Chamber of Commerce (2002). He is founding co-Editor-in-Chief of the new ACM Transactions on Internet of Things (ACM TIoT) as well as Editor-in-Chief of Computing (Springer). He is an Associate Editor of IEEE Transactions on Services Computing, IEEE Transactions on Cloud Computing, ACM Transactions on the Web, and ACM Transactions on Internet Technology, as well as on the editorial board of IEEE Internet Computing and IEEE Computer. Dustdar is recipient of the ACM Distinguished Scientist award (2009), the IBM Faculty Award (2012), an elected member of the Academia Europaea: The Academy of Europe, where he is chairman of the Informatics Section, as well as an IEEE Fellow (2016).

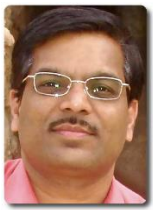

**Rajkumar Buyya** is a Redmond Barry Distinguished Professor and Director of the Cloud Computing and Distributed Systems (CLOUDS) Laboratory at the University of Melbourne, Australia. He is also serving as the founding CEO of Manjrasoft, a spin-off company of the University, commercializing its innovations in Cloud Computing. He served as a Future Fellow of the Australian Research Council during 2012-2016. He has authored over 625 publications and seven text books including "Mastering Cloud Computing" published by McGraw Hill, China Machine Press, and Morgan Kaufmann for Indian, Chinese and international markets respectively. He also edited several books including "Cloud Computing: Principles and Paradigms" (Wiley Press, USA, Feb 2011). He is one of the highly cited authors in computer science and software engineering worldwide (h-index=134, g-index=298, 96,000+ citations). "A Scientometric Analysis of Cloud Computing Literature" by German scientists ranked Dr. Buyya as the World's Top-Cited (#1) Author and the World's Most-Productive (#1) Author in Cloud Computing. Dr. Buyya is recognized as a "Web of Science Highly Cited Researcher" for four consecutive years since 2016, a Fellow of IEEE, and Scopus Researcher of the Year 2017 with Excellence in Innovative Research Award by Elsevier and recently (2019) received "Lifetime Achievement Awards" from two Indian universities for his outstanding contributions to Cloud computing and distributed systems. Software technologies for Grid and Cloud computing developed under Dr. Buyya's leadership have gained rapid acceptance and are in use at several academic institutions and commercial enterprises in 40 countries around the world. Dr. Buyya has led the establishment and development of key community activities, including serving as foundation Chair of the IEEE Technical Committee on Scalable Computing and five IEEE/ACM conferences. These contributions and international research leadership of Dr. Buyya are recognized through the award of "2009 IEEE Medal for Excellence in Scalable Computing" from the IEEE Computer Society TCSC. Manjrasoft's Aneka Cloud technology developed under his leadership has received "2010 Frost & Sullivan New Product Innovation Award". Recently, Dr. Buyya received "Mahatma Gandhi Award" along with Gold Medals for his outstanding and extraordinary achievements in Information Technology field and services rendered to promote greater friendship and India-International cooperation. He served as the founding Editor-in-Chief of the IEEE Transactions on Cloud Computing. He is currently serving as Editor-in-Chief of Journal of Software: Practice and Experience, which was established 50+ years ago. For further information on Dr.Buyya, please visit his cyberhome: www.buyya.com